\RequirePackage{filecontents}

\RequirePackage{snapshot}
\documentclass[aps,pra,reprint,showpacs,secnumarabic,superscriptaddress]{preprint}

\renewcommand\documentclass[2][]{}
\let\origparagraph=\paragraph
\AtBeginDocument{%
  \let\paragraph=\origparagraph
}
\makeatother
\documentclass[aps,pra,reprint,showpacs,floatfix,superscriptaddress]{revtex4-1}

\usepackage[T1]{fontenc}
\usepackage{amsmath}
\usepackage{txfonts}
\usepackage{tgtermes}
\usepackage[altg]{qtxmath}
\renewcommand*{\vec}[1]{\boldsymbol{#1}}
\usepackage{braket}
\usepackage{dsfont}
\usepackage{microtype}
\usepackage{graphicx}
\graphicspath{{figures/}}
\usepackage[breaklinks=true,
pdfauthor={Heiko Bauke, Sven Ahrens, and Rainer Grobe},
pdftitle={Electron-spin dynamics in elliptically polarized light
  waves},pdfborderstyle={/S/D/D[1 0]/W 0.5}]{hyperref}
\usepackage{textcomp}
\usepackage[usenames,dvipsnames]{color}

\newcommand*{\I}{\mathrm{i}}
\newcommand*{\D}{\mathrm{d}}
\newcommand*{\e}{\mathrm{e}}
\newcommand*{\trans}[1]{#1^{\mathsf{T}}}
\newcommand*{\E}{\mathcal{E}}
\DeclareMathOperator{\sgn}{sgn}

\newcommand*{\acommut}[2]{\mathchoice{\left\{#1,#2\right\}}{\{#1,#2\}}{\{#1,#2\}}{\{#1,#2\}}}

\begin{document}

\title{Electron-spin dynamics in elliptically polarized light waves}

\author{Heiko Bauke}
\email{heiko.bauke@mpi-hd.mpg.de}
\affiliation{Max-Planck-Institut f{\"u}r Kernphysik, Saupfercheckweg~1,
  69117~Heidelberg, Germany} 
\author{Sven Ahrens}
\affiliation{Max-Planck-Institut f{\"u}r Kernphysik, Saupfercheckweg~1,
  69117~Heidelberg, Germany} 
\affiliation{{Intense Laser Physics Theory Unit and Department of
    Physics, Illinois State University, Normal, Illinois 61790-4560 USA}} 
\author{Rainer Grobe}
\affiliation{Max-Planck-Institut f{\"u}r Kernphysik, Saupfercheckweg~1,
  69117~Heidelberg, Germany} 
\affiliation{{Intense Laser Physics Theory Unit and Department of
    Physics, Illinois State University, Normal, Illinois 61790-4560 USA}} 

\date{\today}

\begin{abstract}
  We investigate the coupling of the spin angular momentum of light
  beams with elliptical polarization to the spin degree of freedom of
  free electrons.  It is shown that this coupling, which is of similar
  origin as the well-known spin-orbit coupling, can lead to spin
  precession.  The spin-precession frequency is proportional to the
  product of the laser-field's intensity and its spin density.  The
  electron-spin dynamics is analyzed by employing exact numerical
  methods as well as time-dependent perturbation theory based on the
  fully relativistic Dirac equation and on the nonrelativistic Pauli
  equation that is amended by a relativistic correction that accounts
  for the light's spin density.
\end{abstract}


\maketitle

\section{Introduction}
\label{sec:intro}

Novel light sources such as the ELI-Ultra High Field Facility, for
example, envisage to provide field intensities in excess of
$10^{20}\,\mathrm{W/cm^2}$ and field frequencies in the \mbox{x-ray}
domain
\cite{Altarell:etal:2007:XFEL,Yanovsky:Chvykov:etal:2008:Ultra-high_intensity-,McNeil:Thompson:2010:X-ray_free-electron,Emma:Akre:etal:2010:First_lasing,Mourou:Tajima:2011:More_Intense,Mourou:Fisch:etal:2012:Exawatt-Zettawatt_pulse}.
Such facilities open the possibility of probing light-matter interaction
in the ultra-relativistic regime
\cite{Brabec:2008:Strong_Field_Laser_Physics,Ehlotzky:Krajewska:Kaminski:2009:Fundamental_processes,Di-Piazza:Muller:etal:2012:Extremely_high-intensity}.
In this regime, the quantum dynamical interaction of electrons, ions, or
atoms can be modified due to relativistic corrections of the Dirac
equation to the nonrelativistic Schr{\"o}dinger equation.  In certain
setups, however, relativistic effects may change the dynamics not only
quantitatively, but also qualitatively.  As an example, spin-orbit
interaction, that is, the interaction of a particle's spin with its
orbital angular momentum, can lead to a splitting of atomic energy
levels, which are degenerated within the framework of nonrelativistic
quantum theory.  At ultra-high laser intensities also, spin effects can
set in
\cite{Walser:Urbach:etal:2002:Spin_and,Ahrens:Bauke:etal:2012:Spin_Dynamics,Ahrens:Bauke:etal:2013:Kapitza-Dirac_effect}.

In our recent publication
\cite{Bauke:Ahrens:etal:2014:Electron-spin_dynamics}, we demonstrated
that relativistic effects may also modify the spin dynamics of an
electron in external electromagnetic waves with elliptical polarization.
This effect can be attributed to a coupling of the spin density that is
associated with the external laser field to the electron's spin. The
coupling between the electronic spin and the photonic spin is of similar
origin as the well-known spin-orbit effect.  Both effects have in common
that two angular momentum degrees of freedom are coupled that are of
different origin.  Other examples for the interaction between different
kinds of angular momentum have been studied in very different contexts,
e.\,g., the angular momentum of phonons in a magnetic crystal and the
angular momentum of the crystal \cite{Zhang:Niu:2014:Angular_Momentum}
or the orbital angular momentum of light and the internal degrees of
motion in molecules \cite{Mondal:etal:2014:Angular_Momentum_Transfer}, to
mention two recent examples.

In this contribution, we elaborate and extend the theoretical
description of the electron's spin dynamics as studied in
Ref.~\cite{Bauke:Ahrens:etal:2014:Electron-spin_dynamics}.  The article
is organized as follows.  In Sec.~\ref{sec:setup} we describe the
considered laser setup and derive the electron's quantum mechanical
equations of motion, relativistic and nonrelativistic.  Subsequently, we
solve these equations via time-dependent perturbation theory in
Sec.~\ref{sec:perturbation_theory} and by numerical methods in
Sec.~\ref{sec:numerics}.  In Sec.~\ref{sec:fied_strength} we estimate
under which experimental conditions the predicted spin precession may be
realized before we summarize our results in Sec.~\ref{sec:Conclusions}.

\section{Electrons in standing laser fields}
\label{sec:setup}%

\subsection{Laser fields with elliptical polarization}

\begin{figure}
  \centering
  \includegraphics[width=0.95\linewidth]{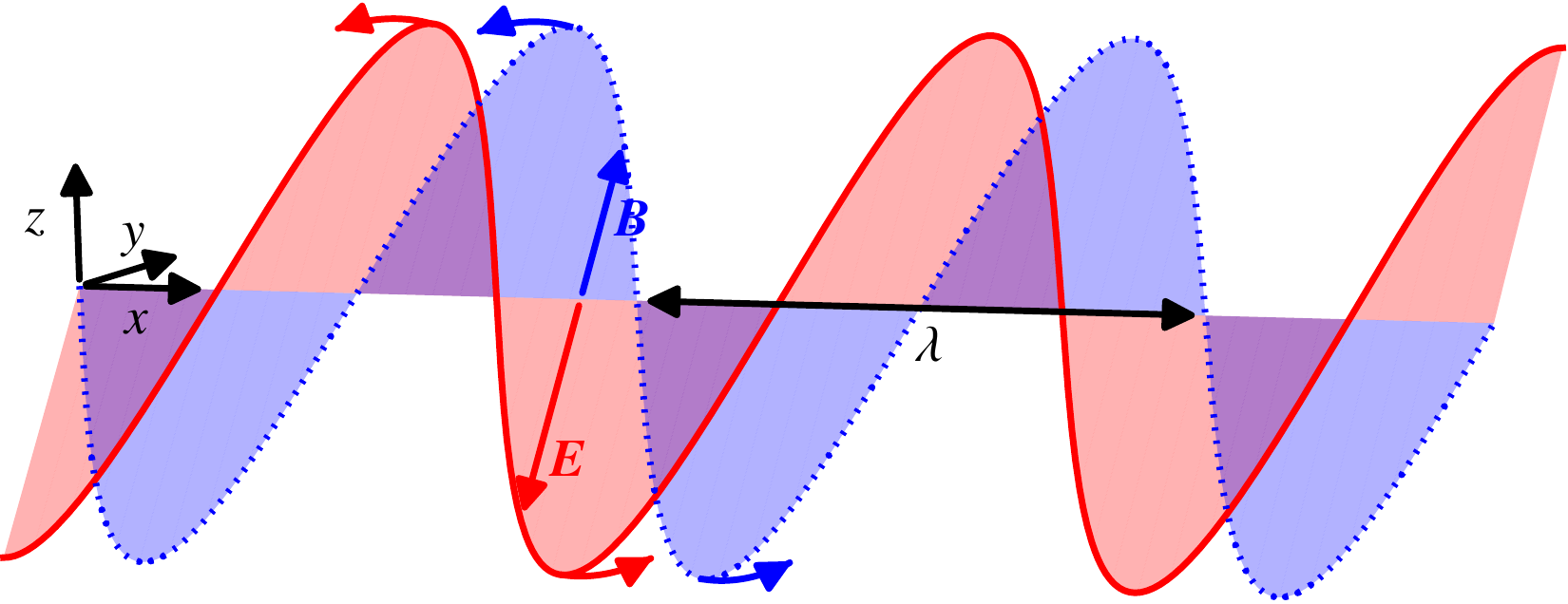}
  \caption{(Color online) Schematic illustration of the electric (solid
    red line) and magnetic (dotted blue line) field components of the
    standing laser wave (\ref{eq:standing_wave}) with $\eta=\pi/2$ that
    is a superposition of two counterpropagating circularly polarized
    laser waves of opposite helicity and equal wavelength $\lambda$.
    The electric and magnetic components $\vec E$ and $\vec B$ are
    parallel to each other and rotate around the propagation direction.}
  \label{fig:laser_setup}
\end{figure}

We examine electrons in standing light waves formed by two
counterpropagating laser beams with circular or elliptic polarization.
The electric and magnetic field components of two elliptically polarized
laser fields counterpropagating in the direction of the $x$~axis are
given by
\begin{subequations}
  \label{eq:elliptic_fields}
  \begin{align}
    \label{eq:elliptic_fields_E}
    \vec E_{1,2}(\vec r, t) & = \hat{E} \left(
      \cos\frac{2\pi(x\mp c t)}{\lambda}\vec e_y + 
      \cos\left(\frac{2\pi(x\mp c t)}{\lambda}\pm\eta\right)\vec e_z
    \right) \,,\\
    \label{eq:elliptic_fields_B}
    \vec B_{1,2}(\vec r, t) & = \frac{\hat{E}}{c} \left(
      \mp\cos\left(\frac{2\pi(x\mp c t)}{\lambda}\pm\eta\right)\vec e_y \pm 
      \cos\frac{2\pi(x\mp c t)}{\lambda}\vec e_z
    \right)\,.
  \end{align}
\end{subequations}
Here, we have introduced the speed of light $c$, the position vector
$\vec r =\trans{(x,y,z)}$, the time $t$, and $\vec e_x$, $\vec e_y$, and
$\vec e_z$ denoting unit vectors in the direction of the coordinate
axes.  The two electromagnetic waves \eqref{eq:elliptic_fields} feature
the same wavelength $\lambda$, the same electric field amplitude
$\hat{E}$, and the same intensity
\begin{equation}
  \label{eq:intensity}
  I = \varepsilon_0c\hat{E}^2 
\end{equation}
but have opposite helicity. The parameter $\eta\in(-\pi,\pi]$ determines
the degree of ellipticity with $\eta=0$ and $\eta=\pi$ corresponding to
linear polarization and $\eta=\pm\pi/2$ to circular polarization.
Introducing the wave number $k=2\pi/\lambda$ and the lasers' angular
frequency $\omega=kc$, the total electric and magnetic fields of the
standing laser wave that is formed by superimposing the two electric and
magnetic field components \eqref{eq:elliptic_fields_E} and
\eqref{eq:elliptic_fields_B} follow as
\begin{subequations}
  \label{eq:standing_wave}
  \begin{align}
    \label{eq:standing_wave_E}
    \vec E(\vec r, t) & = 
    2\hat{E}\cos kx \left( 
      \cos \omega t\,\vec e_y + \cos(\omega t - \eta)\,\vec e_z
    \right) \,,\\
    \label{eq:standing_wave_B}
    \vec B(\vec r, t) & =
    \frac{2\hat{E}}{c}\sin kx \left(
      -\sin(\omega t-\eta)\,\vec e_y + \sin \omega t\,\vec e_z
    \right)\,.
  \end{align}
\end{subequations}
For circular fields ($\eta=\pi/2$), the electric and the magnetic
components (\ref{eq:standing_wave}) are parallel to each other and
rotate around the propagation direction as sketched in
Fig.~\ref{fig:laser_setup}.

The elliptically polarized fields \eqref{eq:elliptic_fields} carry spin
angular momentum
\cite{Beth:1936:Mechanical_Detection,Jackson:1998:Classical_Electrodynamics}.
The partition of a general light beam's total angular momentum into
orbital angular momentum and spin angular momentum was debated in the
literature.  A definition that seems to be well established now
\cite{Stewart:2005:Angular_momentum,Barnett:2011:On_six,Cameron:Barnett:2012:Electric-magnetic_symmetry,Bliokh:Bekshaev:etal:2013:Dual_electromagnetism}
utilizes Coulomb gauge vector potentials for both the magnetic as well
as the electric field component.  The Coulomb gauge vector potentials
$\vec A_{1,2}$ and $\vec C_{1,2}$ of the elliptically polarized
fields~\eqref{eq:elliptic_fields} are
\begin{subequations}
  \label{eq:vector_fields}
  \begin{align}
    \vec C_{1,2}(\vec r, t) & = -\frac{\hat{E}}{\omega} \left(
      \pm\sin(kx \mp \omega t\pm\eta)\,\vec e_y 
      -\sin(kx \mp \omega t)\,\vec e_z
    \right) \,,\\
    \vec A_{1,2}(\vec r, t) & = -\frac{\hat{E}}{\omega} \left(
      \mp\sin(kx \mp \omega t)\,\vec e_y 
      \mp\sin(kx \mp \omega t\pm\eta)\,\vec e_z
    \right) \,.
  \end{align}
\end{subequations}
These potentials are related to the electromagnetic fields via $\vec
E_{1,2}=-\dot{\vec A}_{1,2}= -c\vec\nabla\times\vec C_{1,2}$, %
$\vec B_{1,2}=\vec\nabla\times\vec A_{1,2}=-\dot{\vec C}_{1,2}/c$ and the
potentials comply with the Coulomb gauge as $\vec\nabla\cdot\vec
A_{1,2}=\vec\nabla\cdot\vec C_{1,2}=0$.  With these quantities the spin
density of each laser field is given by the expression
$\varepsilon_0(\vec E_{1,2}\times \vec A_{1,2}+c\vec B_{1,2}\times\vec
C_{1,2})/2$, which reduces in the case of the plane wave fields
\eqref{eq:elliptic_fields} to
\begin{equation}
  \label{eq:spin_density}
  \frac{\varepsilon_0}{2} 
  (\vec E_{1,2}\times \vec A_{1,2}+c\vec B_{1,2}\times\vec C_{1,2}) =
  \varepsilon_0 \vec E_{1,2}\times \vec A_{1,2} =
  \frac{\varepsilon_0 \hat{E}^2\lambda\sin\eta}{2\pi c}\vec{e}_x\,.
\end{equation}
The total photonic spin density of the setup is
\begin{equation}
  \frac{\varepsilon_0}{2} 
  (\vec E\times \vec A+c\vec B\times\vec C) =
  \frac{\varepsilon_0 \hat{E}^2\lambda\sin\eta}{\pi c}\vec{e}_x\,,
\end{equation}
with $\vec E=\vec E_{1}+\vec E_{2}$, $\vec B=\vec B_{1}+\vec B_{2}$,
$\vec A=\vec A_{1}+\vec A_{2}$, and $\vec C=\vec C_{1}+\vec C_{2}$.
Thus, the ellipticity parameter $\eta$ determines also the light's
spin density.

Note, that for plane wave fields the photonic spin density may be
expressed solely by the electric field and the magnetic vector
potential, see \eqref{eq:spin_density}, and many works utilize the
expression \mbox{$\varepsilon_0\vec E\times \vec A$} to define the
photonic spin density, which is not equivalent to $\varepsilon_0(\vec
E\times \vec A+c\vec B\times\vec C)/2$ for none-plane-wave fields,
e.\,g., standing waves.  The total amount of spin angular momentum,
however, does not depend on which of both densities is chosen
\cite{Barnett:2010:Rotation_of}, this means
\begin{equation}
  \int \varepsilon_0\vec E\times \vec A \,\D^3 r =
  \int \frac{\varepsilon_0}{2} 
  (\vec E\times \vec A+c\vec B\times\vec C)\,\D^3 r \,,
\end{equation}  
where the integration goes over all space regions with non-vanishing
electromagnetic fields.

Furthermore, note that the given expressions for the photonic spin
density require that the electromagnetic vector potentials $\vec A$ and
$\vec C$ are given in Coulomb gauge.  If another gauge is applied then
$\vec A$ and $\vec C$ in the definition of the spin density have to be
replaced by $\vec A^\perp$ and $\vec C^\perp$, which are defined by
splitting the vector potentials $\vec A$ and $\vec C$ such that $\vec
A=\vec A^\perp+\vec A^\parallel$ and $\vec\nabla\cdot \vec A^\perp=0$
and $\vec\nabla\times\vec A^\parallel=0$ and similarly for the electric
vector potential $\vec C$ \cite{Cameron:2014:On_second}.  Thus the
Coulomb gauge is the most convenient gauge to study effects of the
photonic spin and it is therefore employed here.  All predicted effects
of this contribution result from gauge invariant quantum mechanical wave
equations and are, therefore, independent of the choice of the gauge.

\subsection{Quantum dynamics in momentum space}

Introducing the window function, which allows for a smooth turn-on and
turn-off of the laser field,
\begin{equation}
  \label{eq:turn_on_and_off}
  w(t) = 
  \begin{cases}
    \sin^2 \frac{\pi t}{2\Delta T} & \text{if $0\le t\le \Delta T$,} \\
    1 & \text{if $\Delta T\le t\le T-\Delta T$,} \\
    \sin^2 \frac{\pi (T-t)}{2\Delta T} & \text{if $T-\Delta T\le t\le T$,} 
  \end{cases}
\end{equation}
the magnetic vector potential of the combined laser fields
\eqref{eq:standing_wave} is given by
\begin{equation}
  \label{eq:A}
  \vec A(\vec r, t) = 
  -\frac{2w(t)\hat{E}}{\omega}\cos kx
  \left( 
    \sin \omega t\,\vec e_y + \sin(\omega t-\eta)\,\vec e_z
  \right)\,.
\end{equation}
The parameters $T$ and $\Delta T$ denote the total interaction time and
the turn-on and turn-off intervals.  The quantum mechanical evolution of
an electron of mass $m$ and charge $q=-e$ in the laser field with the
vector potential \eqref{eq:A} is governed by the Dirac equation
\begin{equation}
  \label{eq:Dirac_equation}
  \I \hbar\dot \Psi(\vec r, t)  = 
  \left(
    c \vec \alpha\cdot\big(
    -\I\hbar\vec\nabla - q \vec A(\vec r, t)
    \big) + m c^2 \beta
  \right) \Psi(\vec r, t)
\end{equation}
with the Dirac matrices
$\vec\alpha=\trans{(\alpha_x,\alpha_y,\alpha_z)}$ and $\beta$
\cite{Gross:2004:Relativistic_Quantum,Thaller:2005:Advanced_Visual}.
The quasi one-dimensional sinusoidal structure of the vector potential
(\ref{eq:A}) allows us to cast the partial differential equation
(\ref{eq:Dirac_equation}) into a set of coupled ordinary differential
equations \cite{Ahrens:Bauke:etal:2013:Kapitza-Dirac_effect}.  For this
purpose, we make the ansatz
\begin{equation}
  \label{eq:Dirac_wave_function}
  \Psi(\vec r, t) = \sum_{n,\gamma} c_n^\gamma(t) \psi_n^\gamma(\vec r)
\end{equation}
with integer $n$, $\gamma\in\{+{\uparrow}, -{\uparrow}, +{\downarrow},
-{\downarrow}\}$, and the four-component basis functions
\begin{equation}
  \label{eq:Dirac_momentum_eigen}
  \psi_n^{\gamma}(\vec r) = 
  \sqrt{\frac{k}{2 \pi}} \,u_n^\gamma \e^{\I n k x}\,,
\end{equation}
with $u_n^\gamma$ defined as
\begin{subequations}
  \label{eq:Dirac_bi-spinors}
  \begin{align}
    u_n^{+\uparrow/+\downarrow} & = \sqrt{\frac{\E_n + m c^2}{2 \E_n}}
    \begin{pmatrix}
      \chi^{\uparrow/\downarrow} \\[0.67ex]
      \frac{n c k \hbar\sigma_x}{\E_n + m c^2} \chi^{\uparrow/\downarrow}
    \end{pmatrix} \,,\\
    u_n^{-\uparrow/-\downarrow} & = \sqrt{\frac{\E_n + m c^2}{2 \E_n}}
    \begin{pmatrix}
      - \frac{n c k \hbar\sigma_x}{\E_n + m c^2} \chi^{\uparrow/\downarrow} \\[0.67ex]
      \chi^{\uparrow/\downarrow}
    \end{pmatrix}\,,
  \end{align}
\end{subequations}
$\chi^\uparrow = \trans{(1, 0)}$, and $\chi^\downarrow = \trans{(0, 1)}$,
and the relativistic energy-momentum relation
\begin{equation}
  \label{eq:relativistic_enery_momentum_relation_modes}
  \E_n = \sqrt{(m c^2)^2 + (n c k\hbar)^2} \,.
\end{equation}
Introducing the four-tuples
\begin{equation}
  c_n(t)=\trans{(c_n^{+\uparrow}(t),c_n^{+\downarrow}(t),c_n^{-\uparrow}(t),c_n^{-\downarrow}(t))}
\end{equation}
and employing the expansion \eqref{eq:Dirac_wave_function} the Dirac
equation becomes in momentum space
\begin{equation}
  \label{eq:Dirac_equation_momentum-space}
  \I \hbar\dot c_n(t) = 
  \E_n \beta c_n(t) + \sum_{n'} V_{n,n'}(t) c_{n'}(t) \,.
\end{equation}
Here, $V_{n,n'}(t)$ denotes the interaction Hamiltonian. Its
components are given by
\begin{multline}
  \label{eq:Dirac_interaction-hamiltonian_comp}
  V_{n,n'}^{\gamma,\gamma'}(t) = 
  \frac{w(t)q\hat E}{k} \left(
    \delta_{n,n'-1}+\delta_{n,n'+1}
  \right) \times\\
  \left(
    {u_n^{\gamma}}^\dagger \alpha_y u_{n'}^{\gamma'} \sin\omega t +
    {u_n^{\gamma}}^\dagger \alpha_z u_{n'}^{\gamma'} \sin(\omega t -\eta)
  \right)\,,
\end{multline}
with $\gamma,\gamma'\in\{+{\uparrow}, -{\uparrow}, +{\downarrow},
-{\downarrow}\}$ and $\delta_{n,n'}$ denoting the Kronecker delta.  As
the basis functions (\ref{eq:Dirac_bi-spinors}) are common
eigenfunctions of the free Dirac Hamiltonian, the canonical momentum
operator, as well as of the $z$~component of the Foldy-Wouthuysen spin
operator \cite{Foldy:Wouthuysen:1950:Non-Relativistic_Limit}, the spin
expectation value (with the $z$~axis as the quantization axis) is
conveniently given by the absolute squared values of the coefficients
$c_n^\gamma(t)$ via
\begin{equation}
  \label{eq:Dirac_spin_exp}
  s_z(t) = \frac{\hbar}{2}
  \sum_n 
  |c_n^{+\uparrow}(t)|^2 + |c_n^{-\uparrow}(t)|^2 -
  |c_n^{+\downarrow}(t)|^2 - |c_n^{-\downarrow}(t)|^2 \,.
\end{equation}

In order to interpret the electron's relativistic quatum dynamics it
will be beneficial to consider the weakly relativistic limit of the
Dirac equation \eqref{eq:Dirac_equation}.  In this limit, this equation
reduces via a Foldy-Wouthuysen transformation
\cite{Foldy:Wouthuysen:1950:Non-Relativistic_Limit,Vries:1970:Foldy-Wouthuysen_Transformations,Frohlich:Studer:1993:Gauge_invariance}
to
\begin{multline}
  \label{eq:Dirac_FW_equation}
  \I \hbar\dot \Psi(\vec r, t)  = 
  \Bigg( 
  \frac{( -\I\hbar\vec\nabla - q \vec A(\vec r, t) )^2}{2m} -
  \frac{q\hbar}{2m}\vec\sigma\cdot \vec B(\vec r, t) 
  +
  q\phi(\vec r, t) \\ 
  -
  \frac{( -\I\hbar\vec\nabla - q \vec A(\vec r, t) )^4}{8m^3c^2} -
  \frac{q^2\hbar^2}{8m^3c^4}(c^2\vec B(\vec r, t)^2-\vec E(\vec r, t)^2) \\
  -
  \frac{q\hbar}{4m^2c^2 }\vec\sigma\cdot (\vec E(\vec r, t) \times (
  -\I\hbar\vec\nabla - q\vec A(\vec r, t) ) ) -
  \frac{q\hbar^2}{8m^2c^2}\vec\nabla\cdot\vec{E}(\vec r, t) \\+
  \frac{q\hbar}{8m^3c^2}\acommut{\vec\sigma\cdot\vec B(\vec r, t)}{( -\I\hbar\vec\nabla - q \vec A(\vec r, t) )^2}
  \Bigg)\,\Psi(\vec r, t) 
\end{multline}
for the now two-component wave function $\Psi(\vec r, t)$ with $\vec
A(\vec r, t)$ given by \eqref{eq:A}, $\vec B(\vec r, t)=\vec\nabla\times\vec
A(\vec r, t)$, $\vec E(\vec r, t)=-\dot{\vec A}(\vec r, t)$, and the
vector of Pauli matrices
$\vec\sigma=\trans{(\sigma_x,\sigma_y,\sigma_z)}$.  In leading order
Eq.~\eqref{eq:Dirac_FW_equation} features four terms that may cause spin
dynamics.  The so-called Zeeman term $\sim\vec\sigma\cdot \vec B(\vec r,
t)$ mediates the coupling of the electron's spin to the magnetic field
and the anticommutator expression
\mbox{$\sim\acommut{\vec\sigma\cdot\vec B(\vec r, t)}{(
    -\I\hbar\vec\nabla - q \vec A(\vec r, t) )^2}$} is the lowest order
relativistic correction to the Zeeman coupling.  It is well-known that
the term $\sim\vec\sigma\cdot\vec E(\vec r, t) \times \I\hbar\vec\nabla$ in
\eqref{eq:Dirac_FW_equation} leads to the so-called spin-orbit
interaction, i.\,e., the coupling between the electron's spin and its
orbital angular momentum.  The term $\sim\vec\sigma\cdot(\vec E(\vec r,
t) \times \vec A(\vec r, t))$, however, may be interpreted as a coupling
of the spin density of the external electromagnetic wave to the
particle's spin.  In Sec.~\ref{sec:numerics} we will show that the
relativistic electron motion in the setup of Sec.~\ref{sec:setup} can be
described in leading order by the nonrelativistic Pauli equation
complemented by the relativistic correction due to the electromagnetic
wave's spin density as the only relativistic correction.  This leads us
to the relativistic Pauli equation
\begin{multline}
  \label{eq:Pauli_FW_equation}
  \I \hbar\dot \Psi(\vec r, t)  = 
  \Bigg(
  \frac{1}{2m}( -\I\hbar\vec\nabla - q \vec A(\vec r, t) )^2 -
  \frac{q\hbar}{2m}\vec\sigma\cdot \vec B(\vec r, t) \\ +
  \frac{q^2\hbar}{4m^2c^2 }\vec\sigma\cdot (\vec E(\vec r, t) \times \vec A(\vec r, t) )
  \Bigg)\,\Psi(\vec r, t) \,.
\end{multline}

We also formulate the relativistic Pauli equation
\eqref{eq:Pauli_FW_equation} in momentum space by utilizing the ansatz
\eqref{eq:Dirac_wave_function} as for the Dirac equation but now with
$\gamma\in\{\uparrow, \downarrow\}$ and the two-component basis
functions
\begin{equation}
  \label{eq:Pauli_momentum_eigen}
  \psi_n^{\gamma}(\vec r) = 
  \sqrt{\frac{k}{2 \pi}} \,\chi^\gamma \e^{\I n k x}\,.
\end{equation}
Using the expansion \eqref{eq:Dirac_wave_function} with the functions
\eqref{eq:Pauli_momentum_eigen} and introducing the pair
\begin{equation}
  c_n(t)=\trans{(c_n^{\uparrow}(t),c_n^{\downarrow}(t))} \,,
\end{equation}
the relativistic Pauli \eqref{eq:Pauli_FW_equation} is given in momentum
space by
\begin{multline}
  \label{eq:Pauli_FW_equation_momentum-space}
  \I \hbar\dot c_n(t) = 
  \frac{n^2k^2\hbar^2}{2 m}c_n(t) \\ + 
  \frac{q^2{\hat E}^2 w(t)^2}{2k^2 m c^2}(1-\cos\eta\cos(2\omega t-\eta))
  (c_{n-2}(t)+2c_{n}(t)+c_{n+2}(t)) \\ +
  \frac{\I \hbar q \hat E w(t)}{2mc} 
  (-\sigma_y\sin(\omega t-\eta) + \sigma_z\sin \omega t)
  (c_{n-1}(t)-c_{n+1}(t)) \\ +
  \frac{\hbar q^2 {\hat E}^2 w(t)^2\sin\eta}{4km^2 c^3}\sigma_x 
  (c_{n-2}(t)+2c_{n}(t)+c_{n+2}(t)) \,.
\end{multline}
Neglecting also the relativistic corrections due to the term
$\sim\vec\sigma\cdot(\vec E(\vec r, t) \times \vec A(\vec r, t))$ in
\eqref{eq:Pauli_FW_equation},
Eq.~\eqref{eq:Pauli_FW_equation_momentum-space} reduces to the
nonrelativistic Pauli equation in momentum space
\begin{multline}
  \label{eq:Pauli_equation_momentum-space}
  \I \hbar\dot c_n(t) = 
  \frac{n^2k^2\hbar^2}{2 m}c_n(t) \\ + 
  \frac{q^2{\hat E}^2 w(t)^2}{2k^2 m c^2}(1-\cos\eta\cos(2\omega t-\eta))
  (c_{n-2}(t)+2c_{n}(t)+c_{n+2}(t)) \\ +
  \frac{\I \hbar q \hat E w(t)}{2mc} 
  (-\sigma_y\sin(\omega t-\eta) + \sigma_z\sin \omega t)
  (c_{n-1}(t)-c_{n+1}(t)) \,.
\end{multline}
The two-component basis functions \eqref{eq:Pauli_momentum_eigen} are
common eigenfunctions of the free Pauli Hamiltonian, the canonical
momentum operator, and the nonrelativistic spin operator
$\hbar\sigma_z/2$. Thus, the spin expectation value (with the $z$~axis
as the quantization axis) is given by the absolute squared values of the
coefficients $c_n^\gamma(t)$ via
\begin{equation}
  \label{eq:Pauli_spin_exp}
  s_z(t) = \frac{\hbar}{2}
  \sum_n 
  |c_n^{\uparrow}(t)|^2 -
  |c_n^{\downarrow}(t)|^2 \,.
\end{equation}

Equations \eqref{eq:Dirac_equation_momentum-space},
\eqref{eq:Pauli_FW_equation_momentum-space}, and
\eqref{eq:Pauli_equation_momentum-space} are the basic equations that we
are going to solve in the next sections numerically and analytically via
time-dependent perturbation theory.  In the following, we assume that
the electron is initially at rest with spin aligned to the $z$ axis,
which corresponds to the initial condition $c_0^{{+}\uparrow}(0)=1$ (or
$c_0^{\uparrow}(0)=1$ in case of the Pauli equation) and
$c_n^\gamma(0)=0$ else.

\section{Time-dependent perturbation theory}
\label{sec:perturbation_theory}

The short-time evolution of the electron dynamics may be calculated via
time-dependent perturbation theory \cite{Joachain:etal:2011:Atoms} that
allows us to devise an analytic expression for the Rabi frequency of the
spin precession.  In this scheme the Hamiltonian is split into a free
part and a possibly time-dependent interaction part, viz., $H=H_0+V(t)$.
Then the time evolution operator that maps a quantum state from time $0$
to time $t$ is given by
\begin{equation}
  U(t) = U_0(t) + U_1(t) + U_2(t) + \dots 
\end{equation}
with the free propagator
\begin{equation}
  \label{eq:free_dirac_propagator}
  U_0(t) = \exp\left(
    -\frac{\I}{\hbar} H_0 t
  \right)
\end{equation}
and the corrections due to the interaction Hamiltonian $V(t)$
\begin{multline}
  \label{eq:time_ordert_integrals}
  U_n(t) = \frac{1}{(\I\hbar)^n}
  \int_{0}^t\D t_n \dots
  \int_{0}^{t_3}\D t_{2}
  \int_{0}^{t_2}\D t_1 \\
  U_0(t - t_n)V(t_n)\dots 
  U_0(t_2 - t_1)V(t_1)U_0(t_1)
\end{multline}
for $n>0$.  To simplify the following calculations we neglect the
turn-on and turn-off times; i.\,e., we set $\Delta T=0$ for the reminder
of this section.  Furthermore, we specialize to the case of laser waves
with circular polarization, i.\,e., $\eta=\pi/2$.

\subsection{Dirac equation}
\label{sec:perturbation_theory_Dirac}

For the Dirac equation in momentum space
\eqref{eq:Dirac_equation_momentum-space} the operators $H_0$, $V(t)$,
and $U_0(t, 0)$ are represented by matrices of infinite dimension that
we will conveniently write in terms of sub-matrices of size $2\times2$
and $4\times4$.  First, we introduce the $2\times2$ matrices
\begin{subequations}
  \begin{align}
    V_{y;n,n'}^{\zeta,\zeta'} & =
    \begin{pmatrix}
      {u_n^{\zeta,\uparrow}}^\dagger \alpha_y u_{n'}^{\zeta',\uparrow} &
      {u_n^{\zeta,\uparrow}}^\dagger \alpha_y u_{n'}^{\zeta',\downarrow} \\
      {u_n^{\zeta,\downarrow}}^\dagger \alpha_y u_{n'}^{\zeta',\uparrow}
      & {u_n^{\zeta,\downarrow}}^\dagger \alpha_y
      u_{n'}^{\zeta',\downarrow}
    \end{pmatrix} \,,\\
    V_{z;n,n'}^{\zeta,\zeta'} & =
    \begin{pmatrix}
      {u_n^{\zeta,\uparrow}}^\dagger \alpha_z u_{n'}^{\zeta',\uparrow} &
      {u_n^{\zeta,\uparrow}}^\dagger \alpha_z u_{n'}^{\zeta',\downarrow} \\
      {u_n^{\zeta,\downarrow}}^\dagger \alpha_z u_{n'}^{\zeta',\uparrow}
      & {u_n^{\zeta,\downarrow}}^\dagger \alpha_z
      u_{n'}^{\zeta',\downarrow}
    \end{pmatrix} \,,
  \end{align}
  and
  \begin{multline}
    \label{eq:Dirac_interaction-hamiltonian_comp_2x2}
    V_{n,n'}^{\zeta,\zeta'}(t) = \frac{q\hat E}{k} \left(
      \delta_{n,n'-1}+\delta_{n,n'+1}
    \right) \times\\
    \left(
      V_{y;n,n'}^{\zeta,\zeta'} \sin\omega t -
      V_{z;n,n'}^{\zeta,\zeta'} \cos\omega t 
    \right)\,,
  \end{multline}
\end{subequations}
with $\zeta, \zeta'\in\{+,-\}$. We further introduce the $4\times4$
matrix
\begin{equation}
  V_{n,n'}(t) =
  \begin{pmatrix}
    V_{n,n'}^{++}(t) & V_{n,n'}^{+-}(t) \\
    V_{n,n'}^{-+}(t) & V_{n,n'}^{--}(t) 
  \end{pmatrix} 
\end{equation}
that comprises 16 matrix elements
\eqref{eq:Dirac_interaction-hamiltonian_comp} of the interaction
Hamiltonian $V(t)$.  In this notation, the free Hamiltonian $H_0$ reads
\begin{equation}
  H_{0;n,n'} = 
  \begin{pmatrix}
    \E_n \mathds{1} & 0 \\
    0 & -\E_n \mathds{1} 
  \end{pmatrix}\delta_{n,n'}\,.
\end{equation}
Consequently, matrix elements of the free propagator
\eqref{eq:free_dirac_propagator} are given by
\begin{equation}
  \label{eq:free_prop_matrix}
  U_{0;n,n'}(t) =
  \begin{pmatrix}
    U_{0;n,n'}^{++}(t) & 0 \\
    0 & U_{0;n,n'}^{--}(t) 
  \end{pmatrix} 
\end{equation}
with
\begin{subequations}
  \begin{align}
    U_{0;n,n'}^{++}(t) = &
    \exp(- \I \E_n t/\hbar) \delta_{n,n'}\mathds{1} \,,\\
    U_{0;n,n'}^{--}(t) = & 
    \exp(\I \E_n t/\hbar) \delta_{n,n'}\mathds{1} \,.
  \end{align}
\end{subequations}
As our numerical simulations indicate that spin changing quantum
transitions occur from the state with coefficient $c_0^{+\uparrow}$ to
the state with $c_0^{+\downarrow}$, we do not need to calculate the full
time evolution matrix, knowing
\begin{multline}
  U_{0,0}^{++}(t) = 
  U_{0;0,0}^{++}(t) + 
  U_{1;0,0}^{++}(t) + U_{2;0,0}^{++}(t) \\ +
  U_{3;0,0}^{++}(t) + U_{4;0,0}^{++}(t) + 
  \dots
\end{multline}
will be sufficient.  A short calculation shows that $U_{n;0,0}^{++}(t)$
is a zero matrix for odd $n$.  In the follwing we will calculate the
propagators in second and fourth order time-dependent perturbation
theory.

First, we calculate the second-order propagator, that is, according to
\eqref{eq:time_ordert_integrals},
\begin{multline}
  \label{eq:propagator_U200}
  U_{2;0,0}^{++}(t) = \\
  -\frac{1}{\hbar^2}
  \int_{0}^t\D t_{2} \!
  \int_{0}^{t_{2}}\D t_1 \,
  U_{0;0,0}^{++}(t - t_2) 
  \Big( 
   V_{0,1}^{++}(t_2) U_{0;1,1}^{++}(t_2 - t_1) V_{1,0}^{++}(t_1) \\ + 
   V_{0,1}^{+-}(t_2) U_{0;1,1}^{--}(t_2 - t_1) V_{1,0}^{-+}(t_1) \\ +
   V_{0,-1}^{++}(t_2) U_{0;-1,-1}^{++}(t_2 - t_1) V_{-1,0}^{++}(t_1) \\ +
   V_{0,-1}^{+-}(t_2) U_{0;-1,-1}^{--}(t_2 - t_1) V_{-1,0}^{-+}(t_1) 
  \Big)U_{0;0,0}^{++}(t_1) \,.
\end{multline}
Note that in the last equation we took advantage of the fact that the
interaction Hamiltonian and the free particle propagator are banded
matrices.  Rewriting the matrix elements
\eqref{eq:Dirac_interaction-hamiltonian_comp_2x2} of the interaction
Hamiltonian by expanding the trigonometric functions into exponentials
\begin{multline}
  \label{eq:Dirac_interaction-hamiltonian_comp_2x2_exp}
  V_{n,n'}^{\zeta,\zeta'}(t) = \frac{q\hat E}{2k} \left(
      \delta_{n,n'-1}+\delta_{n,n'+1}
    \right) \times\\
    \left(
      \left(
        -\I V_{y;n,n'}^{\zeta,\zeta'} - V_{z;n,n'}^{\zeta,\zeta'}
      \right)\e^{\I\omega t} +
      \left(
        \I V_{y;n,n'}^{\zeta,\zeta'} - V_{z;n,n'}^{\zeta,\zeta'}
      \right)\e^{-\I\omega t}
    \right)\,,
\end{multline}
and utilizing the identity
\begin{multline}
  \label{eq:spinor_contraction}
  \begin{pmatrix}
    \pm\I V_{y;n,n'}^{++} - V_{z;n,n'}^{++} & \pm\I V_{y;n,n'}^{+-} - V_{z;n,n'}^{+-} \\
    \pm\I V_{y;n,n'}^{-+} - V_{z;n,n'}^{-+} & \pm\I V_{y;n,n'}^{--} - V_{z;n,n'}^{--} \\
  \end{pmatrix} = \\
  \begin{pmatrix}
    r_{n,n'} (\mp \sigma_z + \I \sigma_y) & t_{n,n'} (\pm \I \sigma_y - \sigma_z) \\
    t_{n,n'} (\pm \I \sigma_y - \sigma_z) & r_{n,n'} (\pm \sigma_z - \I \sigma_y) 
  \end{pmatrix}
\end{multline}
with the coefficients
\begin{align}
  \label{eq:dirac-spinor-coefficient-t}
  t_{n,n'} & = d^+_n d^+_{n'} + d^-_n d^-_{n'}\,, \\
  \label{eq:dirac-spinor-coefficient-r}
  r_{n,n'} & = d^-_n d^+_{n'} - d^+_n d^-_{n'} \,,
\end{align}
and\pagebreak[1]
\begin{subequations}
  \begin{align}
    d_n^+ & = \frac{1}{\sqrt{2}}\sqrt{1+\frac{mc^2}{\E_n}}\,,\\
    d_n^- & = \frac{\sgn n}{\sqrt{2}}\sqrt{1-\frac{mc^2}{\E_n}}
  \end{align}
\end{subequations}
the integral \eqref{eq:propagator_U200} evaluates to
\cite{Ahrens:2012:PhD_thesis}\pagebreak[1]
\begin{multline}
  \label{eq:dirac-second-order-perturbation-theory_evaluated}
  U_\mathrm{2;0,0}^{++}(t) = 
  \I \frac{q^2 {\hat E^2}}{2k^2} \e^{- \I\E_0 t/\hbar}t
  \times\mbox{}\\
  \Bigg(
  \frac{r_{0,1}r_{1,0}(-\mathds{1}+\sigma_x)}{- \E_1 + \E_0 - \hbar\omega} + 
  \frac{r_{0,1}r_{1,0}(-\mathds{1}-\sigma_x)}{- \E_1 + \E_0 + \hbar\omega} +
  \frac{r_{0,-1}r_{-1,0}(-\mathds{1}+\sigma_x)}{- \E_{-1} + \E_0 - \hbar\omega} \\ + 
  \frac{r_{0,-1}r_{-1,0}(-\mathds{1}-\sigma_x)}{-\E_{-1} + \E_0 + \hbar\omega} +
  \frac{t_{0,1}t_{1,0}(\mathds{1}-\sigma_x)}{\E_1 + \E_0 - \hbar\omega} + 
  \frac{t_{0,1}t_{1,0}(\mathds{1}+\sigma_x)}{\E_1 + \E_0 + \hbar\omega} \\ +
  \frac{t_{0,-1}t_{-1,0}(\mathds{1}-\sigma_x)}{\E_{-1} + \E_0 - \hbar\omega} + 
  \frac{t_{0,-1}t_{-1,0}(\mathds{1}+\sigma_x)}{\E_{-1} + \E_0 + \hbar\omega}
  \Bigg)+O(1)\,.
\end{multline}
Note, that we give in
\eqref{eq:dirac-second-order-perturbation-theory_evaluated} only those
terms explicitly that grow linear in time $t$. Terms that oscillate but
remain bounded over time are omitted.  Summing the various terms in
\eqref{eq:dirac-second-order-perturbation-theory_evaluated} finally
yields
\begin{equation}
  \label{eq:dirac-second-order-perturbation-theory_simpified}
  U_\mathrm{2;0,0}^{++}(t) = 
  -\I \frac{q^2 {\hat E^2}}{2k^2 m c^2} \e^{- \I\E_0 t/\hbar}t \,.
\end{equation}
Thus, $U_\mathrm{2;0,0}^{++}(t)$ is diagonal and therefore no spin flip
is predicted in second order time-dependent perturbation
theory.

\begin{widetext}
The fourth order time-dependent perturbation theory propagator is
explicitly given by
\begin{multline}
  \label{eq:propagator_U400}
  U_{4;0,0}^{++}(t) = 
  \frac{1}{\hbar^4}
  \int_{0}^t\D t_{4} \!
  \int_{0}^{t_4}\D t_{3} \!
  \int_{0}^{t_3}\D t_{2} \!
  \int_{0}^{t_2}\D t_1 \,\\
  \sum_{(n_1,n_2,n_3)}
  \sum_{\substack{\zeta_1,\zeta_2,\zeta_3\\\in\{+,-\}}}
  U_{0;0,0}^{++}(t-t_4) 
  V_{0,n_3}^{+,\zeta_3}(t_4) U_{0;n_3,n_3}^{\zeta_3,\zeta_3}(t_4-t_3) 
  V_{n_3,n_2}^{\zeta_3,\zeta_2}(t_3)U_{0;n_2,n_2}^{\zeta_2,\zeta_2}(t_3-t_2)
  V_{n_2,n_1}^{\zeta_2,\zeta_1}(t_2) U_{0;n_1,n_1}^{\zeta_1,\zeta_1}(t_2-t_1) 
  V_{n_1,0}^{\zeta_1+}(t_1) U_{0;0,0}^{++}(t_1) \,,
\end{multline}
where the multi-indices $(n_1,n_2,n_3)$ and ${(\zeta_1,
  \zeta_2,\zeta_3)}$ can take six and eight different values,
respectively.  Thus the sum in \eqref{eq:propagator_U400} runs over 48
terms.  All possible values of the tuple $(n_1,n_2,n_3)$ are shown in
Tab.~\ref{tab:n-index}.  Expanding the trigonometric functions of the
interaction Hamiltonian into exponentials and reordering the terms of
\eqref{eq:propagator_U400} we get
\begin{multline}
  \label{eq:fourth-order-perturbation-theory_expansion}
  U_\mathrm{4;0,0}^{++}(t) = \\
  \left( \frac{q \hat E}{2k\hbar} \right)^4
  \sum_{(n_1,n_2,n_3)} 
  \sum_{\substack{\zeta_1,\zeta_2,\zeta_3\\\in\{+,-\}}}
  \sum_{\substack{\eta_1,\eta_2,\eta_3,\eta_4\\\in\{+1,-1\}}}
  \left( - \eta_4 \I V_{y,0,n_3}^{+,\zeta_3} - V_{z,0,n_3}^{+,\zeta_3} \right)
  \left( - \eta_3 \I V_{y,n_3,n_2}^{\zeta_3,\zeta_2} - V_{z,n_3,n_2}^{\zeta_3,\zeta_2} \right)
  \left( - \eta_2 \I V_{y,n_2,n_1}^{\zeta_2,\zeta_1} - V_{z,n_2,n_1}^{\zeta_2,\zeta_1} \right)
  \left( - \eta_1 \I V_{y,n_1,0}^{\zeta_1,0} - V_{z,n_1,0}^{\zeta_1,+} \right)
  \nu(t)
\end{multline}
with the phase integral
\begin{multline}
  \label{eq:fourth-order-perturbation-theory_phase}
  \nu(t) = \int_{0}^{t} \D t_4 \int_{0}^{t_4} \D t_3 \int_{0}^{t_3} \D
  t_2 \int_{0}^{t_2} \D t_1 \\
  \exp \left( -\frac{\I}{\hbar} 
    \left ( \E_0 (t - t_4) + \zeta_3 \E_{n_3} (t_4 - t_3) + \zeta_2
      \E_{n_2} (t_3 - t_2) + \zeta_1 \E_{n_1} (t_2 - t_1) + \E_0 t_1
    \right) \right)
  \exp \left( 
    \I \omega \left( 
      \eta_4 t_4 + \eta_3 t_3 + \eta_2 t_2 + \eta_1 t_1
    \right) 
  \right)\,.
\end{multline}
\end{widetext}
For those terms where the indices $\eta_1,\eta_2,\eta_3,\eta_4$ sum to
zero the phase integral grows linearly in time, viz.\pagebreak[1]
\begin{multline}
  \nu(t) = 
  \frac{\I}{(\E_0 - \zeta_1 \E_{n_1})/\hbar - \eta_1 \omega}
  \frac{\I}{(\E_0 - \zeta_2 \E_{n_2})/\hbar - (\eta_1 + \eta_2) \omega} \\
  \times
  \frac{\I}{(\E_0 - \zeta_3 \E_{n_3})/\hbar - (\eta_1 + \eta_2 + \eta_3) \omega} 
  \e^{-{\I}\E_0 t/{\hbar} } t + O(1)\,,
\end{multline}
where $O(1)$ indicates some integration constant. If
$\eta_1+\eta_2+\eta_3+\eta_4\ne0$, the function $\nu(t)$ shows
oscillatory behavior, and, therefore, these cases can be neglected.

\begin{table}[t]
  \centering
  \caption{All six possible values of the multi-indices $(n_1,n_2,n_3)$
    that appear in
    the sums of \eqref{eq:propagator_U400} and
    \eqref{eq:fourth-order-perturbation-theory_expansion} and in the integral
    \eqref{eq:fourth-order-perturbation-theory_phase}.} 
  \label{tab:n-index}
  \begin{ruledtabular}
    \begin{tabular}{@{}lccc@{}}
      number & $n_1$ & $n_2$ & $n_3$ \\
      \hline
      1 & 1 & 2 & 1 \\
      2 & 1 & 0 & 1 \\
      3 & 1 & 0 & -1 \\
      4 & -1 & -2 & -1 \\
      5 & -1 & 0 & -1 \\
      6 & -1 & 0 & 1
    \end{tabular}
  \end{ruledtabular}
\end{table}%

A Taylor expansion of the propagator
\eqref{eq:fourth-order-perturbation-theory_expansion} up to the fourth
order in the laser's wave number $k$ yields finally
\begin{equation}
  \label{eq:dirac-fourth-order-perturbation-theory_simpified}
  U_\mathrm{4;0,0}^{++}(t) = 
  \I \left( 
    \Omega_\varphi \mathds{1} + \frac{\Omega}{2} \sigma_{x} 
  \right) \e^{- \I \E_0 t/\hbar}t + O(k^4)\,,
\end{equation}
with the two angular frequencies
\begin{equation}
  \Omega_\varphi = 
  \frac{q^4\hat{E}^4\lambda^6}{(2\pi)^6\hbar^3mc^4}
\end{equation}
and 
\begin{equation}
  \label{eq:Omega_Dirac}
  \Omega = 
  \frac{q^4\hat{E}^4\lambda^5}{(2\pi)^5\hbar^2m^2c^5}\,.
\end{equation}
In contrast to the second order perturbation theory we have found now a
contribution in the propagator that causes a spin precession around the
$x$~axis due to the $\sigma_x$ term in
\eqref{eq:dirac-fourth-order-perturbation-theory_simpified}.  The fourth
order short-time propagator yields for the initial condition
$c_0^{{+}\uparrow}(0)=1$ and $c_n^\gamma(0)=0$ else the spin-flip
probability
\begin{equation}
  \label{eq:c0+down_short}
  |c_0^{+\downarrow}(t)|^2 = \frac{\Omega^2 t^2}{4}\,.
\end{equation}
Anticipating the oscillatory behavior of the long-time spin dynamics
(see Sec.~\ref{sec:numerics} and
Ref.~\cite{Bauke:Ahrens:etal:2014:Electron-spin_dynamics}), which means
\begin{equation}
  \label{eq:c0+down_long}
  |c_0^{+\downarrow}(t)|^2 = \sin^2 \frac{\Omega t}{2}\,,
\end{equation}
we identify the frequency $\Omega$ as the precession frequency
\eqref{eq:Omega_Dirac} of the electron's spin
\cite{Ahrens:Bauke:etal:2013:Kapitza-Dirac_effect} by comparing the
Taylor expansion of \eqref{eq:c0+down_long} with
\eqref{eq:c0+down_short}.  The spin precession frequency
\eqref{eq:Omega_Dirac} is proportional to the photonic spin density
$\varrho_\sigma={\varepsilon_0 \hat{E}^2\lambda}/{(\pi c)}$, the
laser field's intensity $I$ given in \eqref{eq:intensity}, and the
fourth power of the wavelength
\begin{equation}
  \Omega = \varrho_\sigma I \lambda^4 
  \frac{\alpha_\mathrm{el}^2}{2\pi^2m^2 c^3}\,,
\end{equation}
where the proportionality factor ${\alpha_\mathrm{el}^2}/{(2\pi^2m^2
  c^3)}$ is independent of the laser's electromagnetic field and
$\alpha_\mathrm{el}$ denotes the fine-structure constant.

\subsection{Nonrelativistic Pauli equation}
\label{sec:perturbation_theory_Pauli}

In order to calculate the precession frequency in the framework of the
nonrelativistic Pauli equation \eqref{eq:Pauli_equation_momentum-space}
we define the ``free'' Hamiltonian
\begin{equation}
  \label{eq:Pauli_H0}
  H_{0;n,n'} = 
  \left(
    \frac{n^2k^2\hbar^2}{2m} + \frac{q^2{\hat E}^2}{k^2mc^2}
  \right) \delta_{n, n'}\mathds{1}\,.
\end{equation}
Then the interaction Hamiltonian becomes 
\begin{subequations}
  \begin{align}
    V_{n,n'}(t) & = V_{1;n,n'}(t) + V_{2;n,n'}(t) \\
    \intertext{with}
    V_{1;n,n'}(t) & = \frac{\I\hbar q{\hat E}}{2mc}(\sigma_y\cos\omega t -
    \sigma_z\sin\omega t) (\delta_{n, n'+1} - \delta_{n, n'-1}) \,,\\
    V_{2;n,n'}(t) & = \frac{q^2{\hat E}^2}{2k^2mc^2}
    (\delta_{n, n'-2} + \delta_{n, n'+2})\mathds{1} \,.
  \end{align}
\end{subequations}
Neglecting the constant term ${q^2{\hat E}^2}/({k^2mc^2})$ in
\eqref{eq:Pauli_H0}, which causes just a global phase rotation that may
be removed by a suitable gauge, yields the free propagator
\begin{equation}
  U_{0;n,n'}(t) =
  \exp\left(
    - \I  \frac{n^2k^2\hbar}{2m} t
  \right) \delta_{n,n'}\mathds{1} \,.
\end{equation}
The second order propagator for the coefficient $c_0(t)$
\begin{multline}
  \label{eq:Pauli_propagator_U200}
  U_{2;0,0}(t) = \\
  -\frac{1}{\hbar^2}
  \int_{0}^t\D t_{2} \!
  \int_{0}^{t_{2}}\D t_1 \,
  U_{0;0,0}(t-t_2) 
  \Big( 
   V_{2;0,2}(t_2) U_{0;2,2}(t_2-t_1) V_{2;2,0}(t_1) \\ +
   V_{2;0,-2}(t_2) U_{0;-2,-2}(t_2-t_1) V_{2;-2,0}(t_1) \\ +
   V_{1;0,1}(t_2) U_{0;1,1}(t_2-t_1) V_{1;1,0}(t_1) \\ +
   V_{1;0,-1}(t_2) U_{0;-1,-1}(t_2-t_1) V_{1;-1,0}(t_1)
  \Big)U_{0;0,0}(t_1) 
\end{multline}
becomes after time integration
\begin{equation}
  U_{2;0,0}(t) = 
  -\I 
  \frac{q^2\hat E^2}{2m^2c^2}
  \frac{k^2\hbar^2/(2m) \mathds{1} +\omega\hbar\sigma_x}
  {(k^2\hbar^2/(2m))^2 - (\omega\hbar)^2} t +O({\hat E}^4)\,,
\end{equation}
which reduces in the limit $k<mc\hbar$ to
\begin{equation}
  U_{2;0,0}(t) = 
  \I\left(
    \frac{\Omega_\mathrm{P} \hbar k}{4mc}\mathds{1} + 
    \frac{\Omega_\mathrm{P}}{2}\sigma_x
  \right) t \,,
\end{equation}
where 
\begin{equation}
  \label{eq:Omega_P}
  \Omega_\mathrm{P}=\frac{q^2\hat E^2\lambda}{2\pi m^2c^3}
\end{equation}
denotes the spin-precession angular frequency for the nonrelativistic
Pauli equation.  Note that in the case of the nonrelativistic Pauli
equation already second order time-dependent perturbation theory
predicts spin precession in contrast to the fully relativistic Dirac
equation.  Furthermore, spin-precession angular frequencies for the
nonrelativistic Pauli equation \eqref{eq:Omega_P} and the Dirac equation
\eqref{eq:Omega_Dirac} feature qualitatively different dependencies on
the electric field strength $\hat E$: quadratic scaling in the former
case, quartic scaling in the latter case.

\begin{figure}
  \centering
  \includegraphics[scale=0.5]{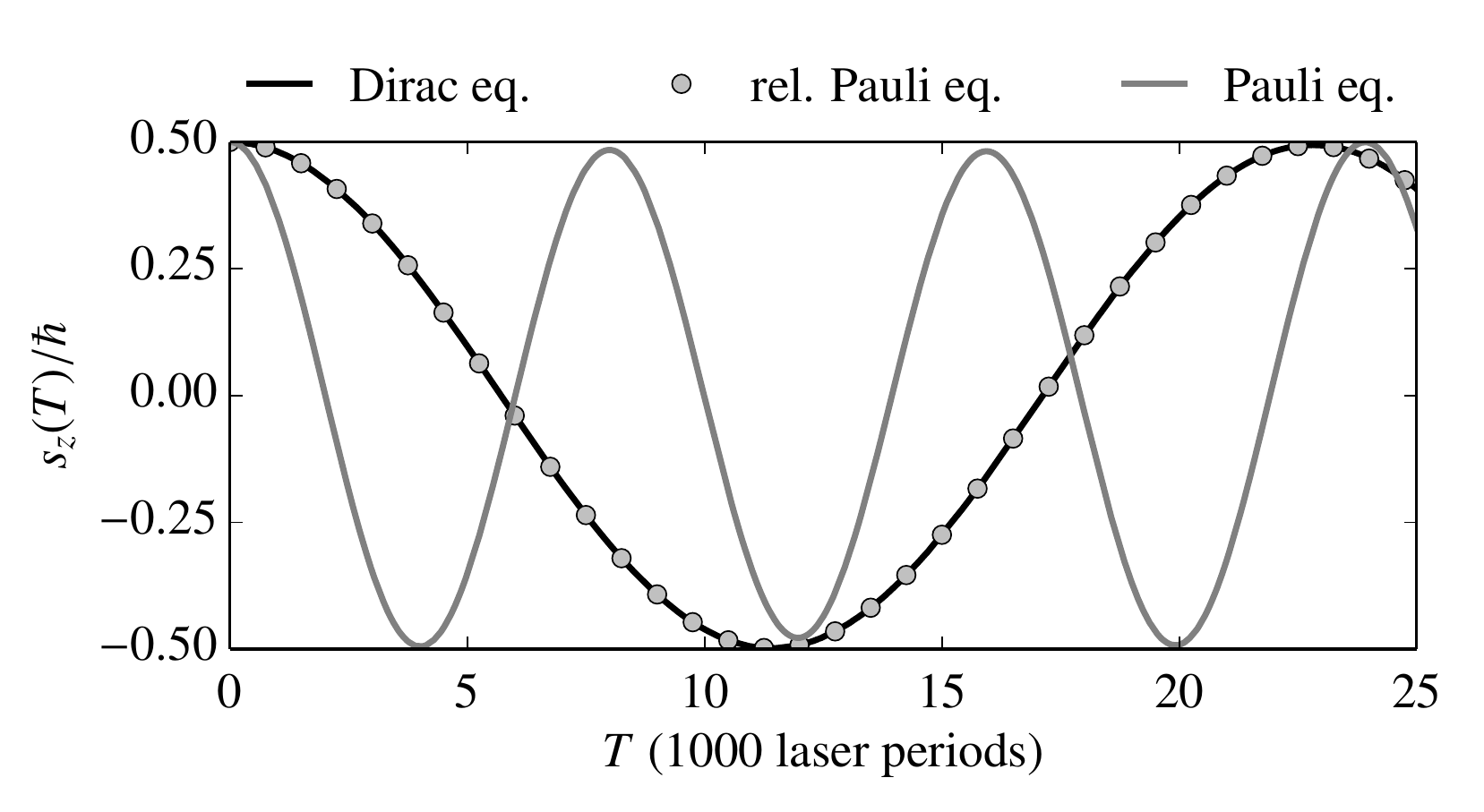}
  \caption{Expectation value of the quantum mechanical electron spin in
    $z$~direction $s_z(T)$ as a function of the total interaction time
    $T$ by using three different equations of motion (the Dirac equation
    \eqref{eq:Dirac_equation_momentum-space}, the relativistic Pauli
    equation \eqref{eq:Pauli_FW_equation_momentum-space}, and the
    nonrelativistic Pauli equation
    \eqref{eq:Pauli_equation_momentum-space}) to model the quatum
    dynamics.  The wavelength and the amplitude of the applied
    circularly polarized laser fields are $\lambda=0.159\,\mathrm{nm}$
    and $\hat E=2.057\times10^{14}\,\mathrm{V/m}$, respectively, and the
    field's switch-on-off interval $\Delta T$ corresponds to five laser
    cycles.}
  \label{fig:spin_dynamics}
\end{figure}

\section{Numerical results}
\label{sec:numerics}

In order to corroborate our analytical results of
Sec.~\ref{sec:perturbation_theory}, we carried out some numerical
simulations solving the time-dependent Dirac equation as well as the
time-dependent relativistic and nonrelativistic Pauli equations.
Figure~\ref{fig:spin_dynamics} shows the expectation value of the
quantum mechanical electron spin in $z$~direction $s_z(T)$ as a function
of the total interaction time $T$.  Time-dependent perturbation theory
predicts different precession frequencies \eqref{eq:Omega_Dirac} and
\eqref{eq:Omega_P} for the nonrelativistic Pauli equation and for the
Dirac equation, respectively. This is consistent with our numerical
simulations, which yield oscillatory behavior of the spin with different
angular frequencies for the nonrelativistic Pauli equation and for the
Dirac equation; see Fig.~\ref{fig:spin_dynamics}.  Note that the
relativistic Pauli equation~\eqref{eq:Pauli_FW_equation_momentum-space}
and the Dirac equation~\eqref{eq:Dirac_equation_momentum-space} yield
spin evolutions that are virtually indistinguishable. Thus, the
relativistic Pauli equation is an excellent approximation to the fully
relativistic Dirac equation for the setup and the parameters that we
consider here.

\begin{figure}
  \centering
  \includegraphics[scale=0.5]{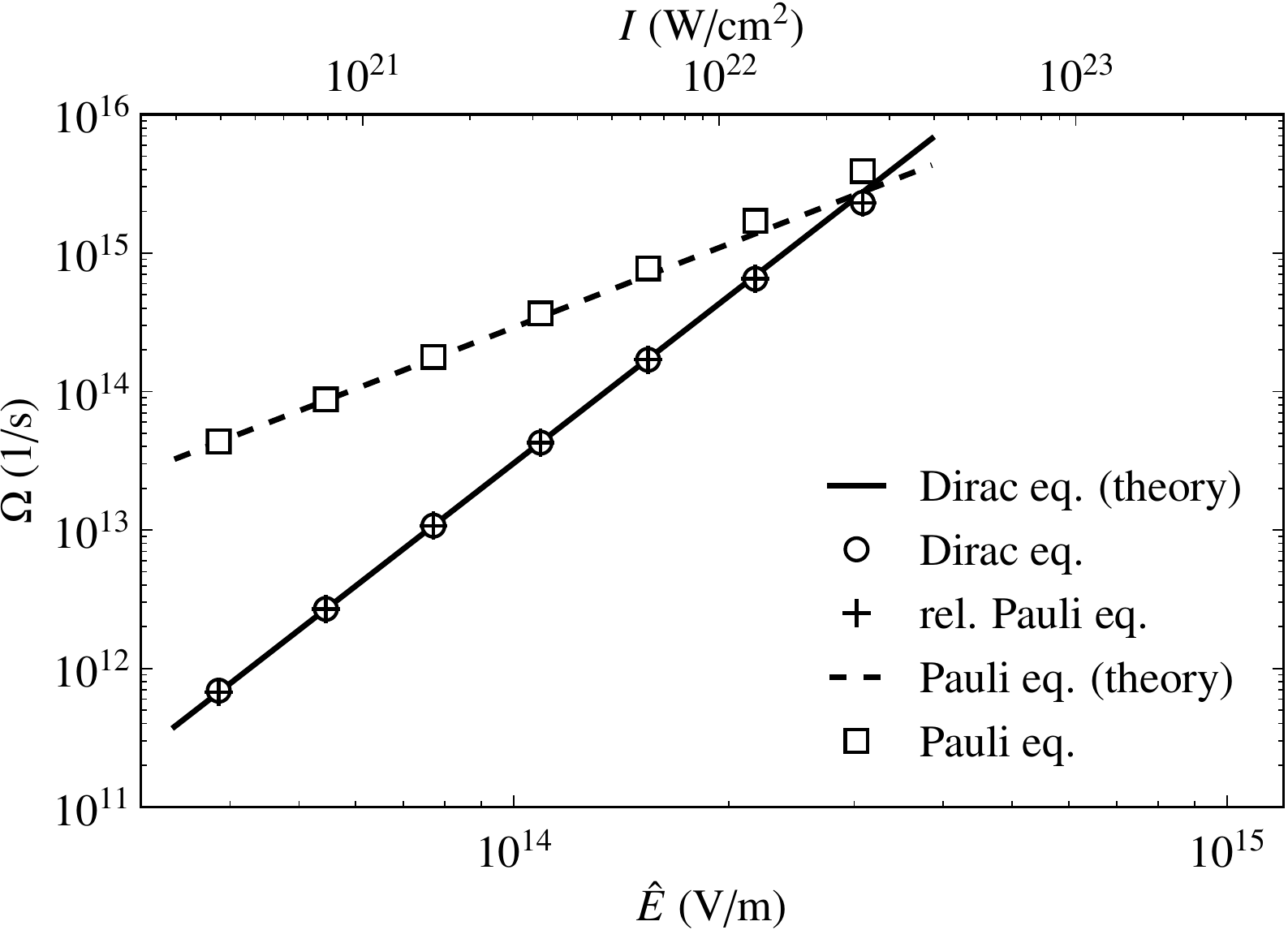}
  \caption{Angular frequency $\Omega$ of the spin precession as a
    function of the laser's electric field strength $\hat{E}$ and its
    intensity $I$ for lasers of the wavelength
    $\lambda=0.159\,\mathrm{nm}$.  Depending on the applied theory (the
    Dirac equation \eqref{eq:Dirac_equation}, the relativistic Pauli
    equation \eqref{eq:Pauli_FW_equation}, or the nonrelativistic Pauli
    equation) the spin of an electron in two counterpropagating
    circularly polarized light waves scales with the second or the
    fourth power of $\hat{E}$.  Theoretical predictions for the
    spin-precession frequency are given by \eqref{eq:Omega_Dirac} and
    \eqref{eq:Omega_P}, respectively.}
  \label{fig:scaling}
\end{figure}

In Fig.~\ref{fig:scaling} we compare the angular frequencies of the spin
precession that result from the nonrelativistic Pauli equation, the
relativistic Pauli equation, and the fully relativistic Dirac equation
for the case of circular polarization ($\eta=\pi/2$).  Our numerical
results reflect the different scaling of the angular frequency as a
function of the peak electric field strength for the nonrelativistic
Pauli equation and the Dirac equation.  The numerical results for the
spin-precession frequency are in an excellent agreement with the
theoretical predictions \eqref{eq:Omega_Dirac} and \eqref{eq:Omega_P},
respectively.  Note that the spin-precession frequency
\eqref{eq:Omega_Dirac} for the Dirac equation also describes the
spin-precession frequency that results from the relativistic Pauli
equation very well.  As demonstrated in
Ref.~\cite{Bauke:Ahrens:etal:2014:Electron-spin_dynamics} the spin
precession angular frequency becomes proportional to $\sin\eta$ if the
light has elliptical polarization.  This means that the spin precession
effect ceases to exist in the case of linear polarization ($\eta=0$).

\begin{figure*}
  \centering
  \includegraphics[scale=0.5]{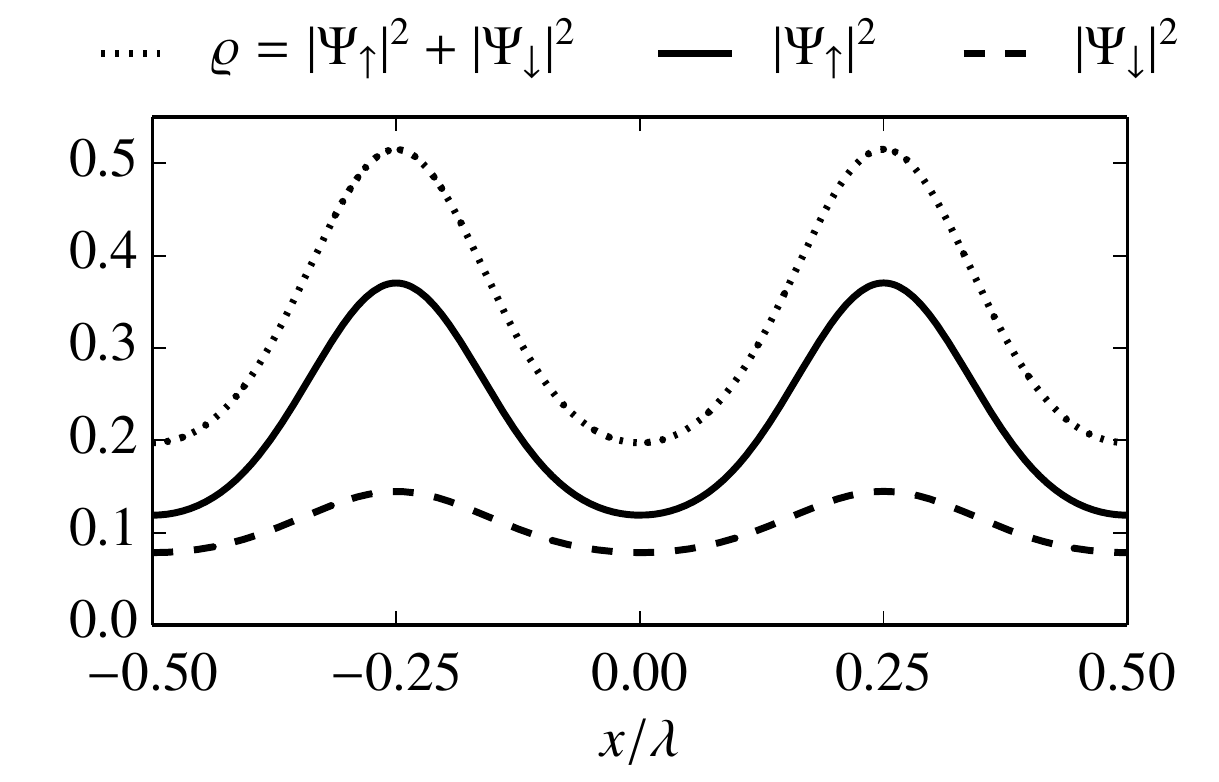}\quad
  \includegraphics[scale=0.5]{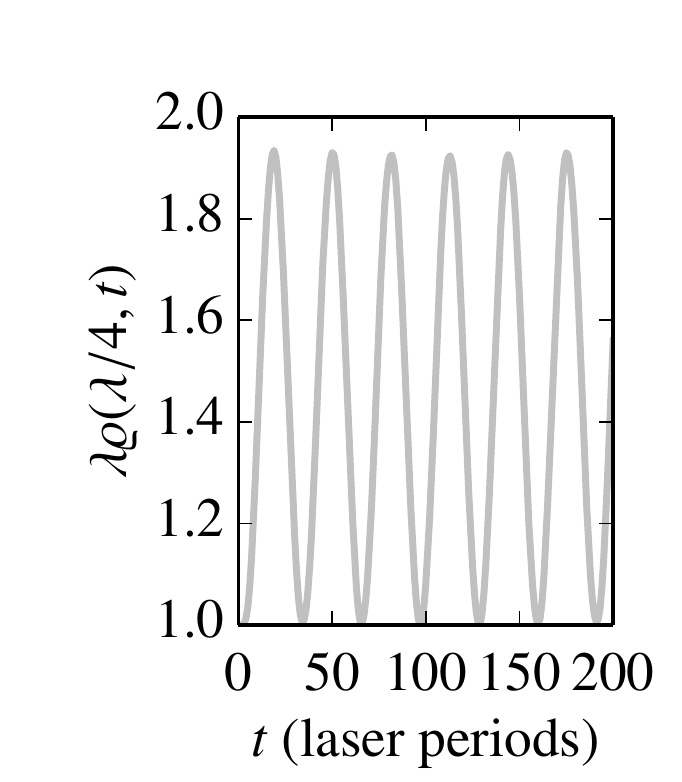}%
  \includegraphics[scale=0.5]{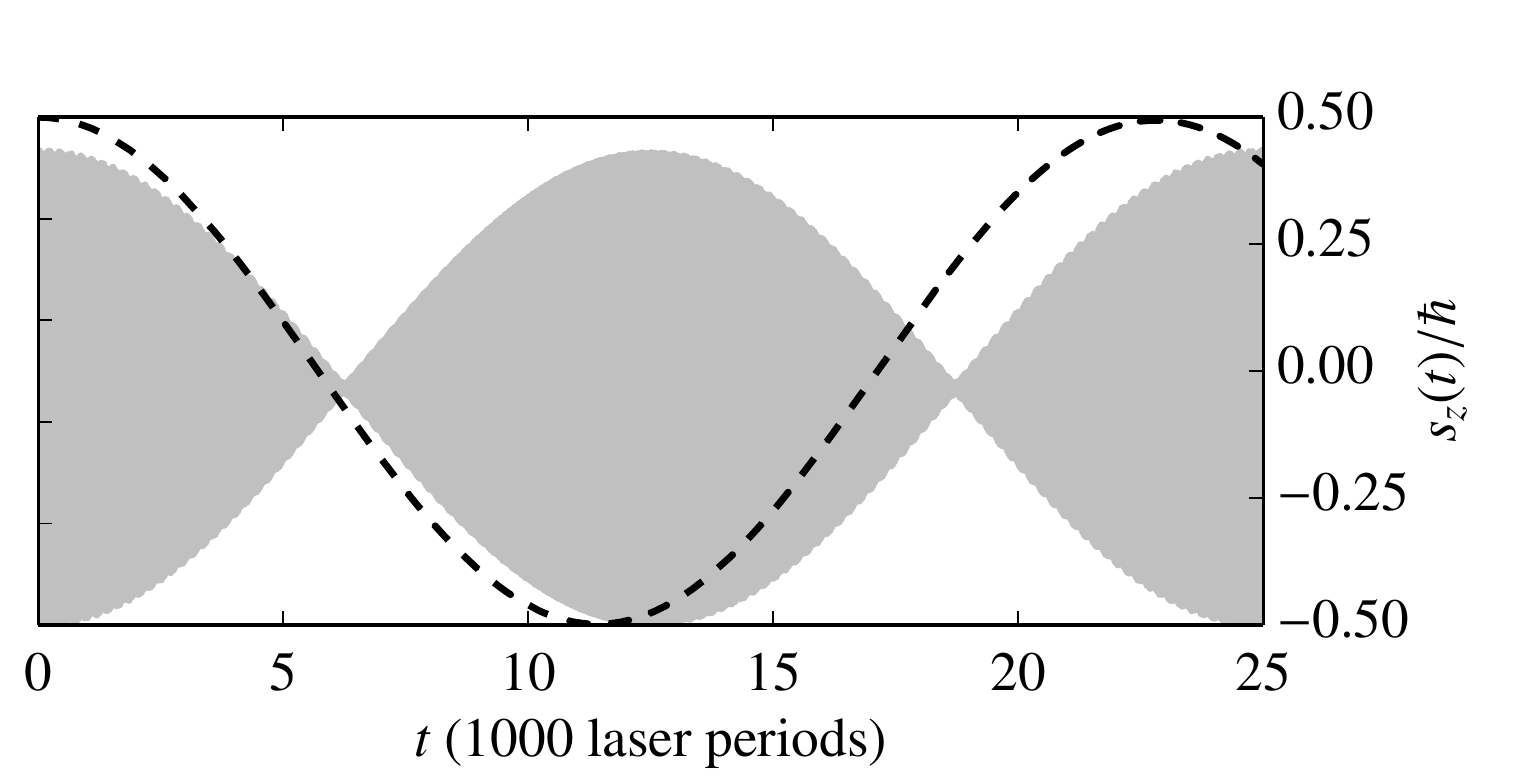}
  \caption{Left display: Electron density
    $\varrho=|\Psi_\uparrow|^2+|\Psi_\downarrow|^2$ and densities of the
    individual spinor components $|\Psi_\uparrow|^2$ and
    $|\Psi_\downarrow|^2$, respectively, after $t=4400$ laser periods.
    The densities accumulate in regions of high magnetic fields near
    $x=\pm\lambda/4$.  Right displays: The scaled electron density
    $\lambda\varrho(x, t)$ at $x=\lambda/4$ (solid gray line, left axis)
    and the expectation value of the spin $s_z(t)$ (dashed black line,
    right axis) as a function of time $t$.  
    All laser parameters as in Fig.~\ref{fig:spin_dynamics}.  Note that
    oscillations of $\varrho(\lambda/4, t)$ are too fast to be resolved
    on the scale of the right most display.}
  \label{fig:wave_packet}
\end{figure*}

In Sec.~\ref{sec:perturbation_theory_Dirac} we demonstrated that in
second order time-dependent perturbation theory for the Dirac equation
there is no spin precession.  In order to derive the spin-precession
frequency we had to calculate the dynamics in fourth order perturbation
theory.  In second order perturbation theory for the relativistic Pauli
theory the spin forces due to the magnetic field and the spin forces due
to the light's spin density just cancel each other.  As shown in
Ref.~\cite{Bauke:Ahrens:etal:2014:Electron-spin_dynamics}, the
cancellation of the spin forces can also be understood by
quasi-classical considerations if one assumes that the electron is
evenly spread over the range of a wavelength.  The observed spin
precession results because the electron's wave function actually
accumulates at the maxima of the magnetic field of the standing light
wave at $x=\pm\lambda/4$.  In order to demonstrate this effect let us
utilize the relativistic Pauli equation \eqref{eq:Pauli_FW_equation}.  A
snapshot of the electron density $\varrho(x, t)=|\Psi_\uparrow(x,
t)|^2+|\Psi_\downarrow(x, t)|^2$ during the electron's interaction with
the laser field is shown in the left part of Fig.~\ref{fig:wave_packet};
here $\Psi_\uparrow$ and $\Psi_\downarrow$ denote the upper and lower
components of the wave function $\Psi$.  The shape of the wave function
can be characterized by calculating the density at $x=\lambda/4$.  If
the density $\varrho(x, t)$ is evenly distributed, then
$\lambda\varrho(\lambda/4, t)=1$. Thus, larger values indicate
deviations from a flat distribution.  As shown in the right part of
Fig.~\ref{fig:wave_packet}, the electron density at $x=\lambda/4$
oscillates at frequencies much larger than the spin-precession
frequency.  The quantity $\lambda\varrho(\lambda/4, t)$ oscillates
around some mean value that is larger than one and that depends on the
electric field amplitude $\hat E$.  Our numerical simulations indicate
that the larger $\hat E$ the more the electron density tends to
concentrate around $x=\pm\lambda/4$. Furthermore, the wave packet shows
a beating behavior as displayed in the right most part of
Fig.~\ref{fig:wave_packet}.  Note that the beat's envelope does not have
the same period as the spin precession.  Thus, the electron's spatial
density and its spin oscillate in an asynchronous manner.

\section{Considerations on the electric field strength}
\label{sec:fied_strength}

\begin{figure}[b]
  \centering
  \includegraphics[scale=0.5]{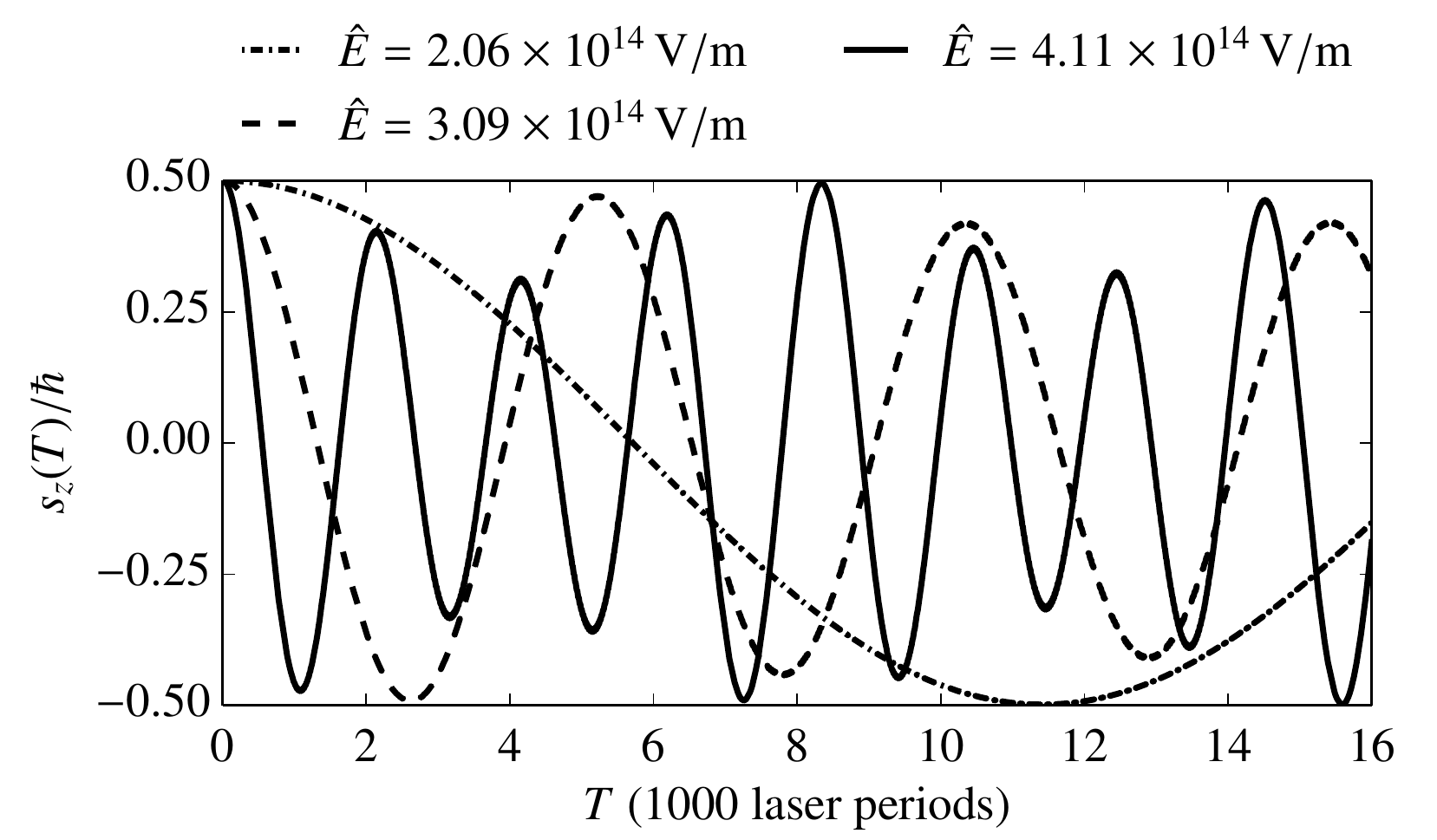}
  \caption{Expectation value of the quantum mechanical electron spin in
    $z$~direction $s_z(T)$ as a function of the total interaction time
    $T$ for different peak electric field strengths $\hat E$.  Other
    parameters as in Fig.~\ref{fig:spin_dynamics}.  The quantum dynamics
    was simulated by solving the time-dependent Dirac equation.}
  \label{fig:spin_dynamics_E}
\end{figure}

The sinusoidal oscillatory motion of the electron's spin in the standing
laser field of two elliptically polarized laser fields is of
perturbative character.  It starts to breakdown if the laser field
becomes too strong resulting in an anharmonic motion of the spin.
Criteria for the breakdown of the harmonic spin precession may be
devised from the relativistic Pauli equation
\eqref{eq:Pauli_FW_equation_momentum-space}. The external laser field
can be treated as a small perturbation if the kinetic-energy coefficient
${k^2\hbar^2}/{(2m)}$ in \eqref{eq:Pauli_FW_equation_momentum-space} is
large compared to the others. This means the conditions
\begin{equation}
  \label{eq:perturbative_cond}
  \frac{q^2{\hat E}^2}{2k^2 mc^2}  < \frac{k^2\hbar^2}{2m}\,, \quad
  \frac{\hbar |q|\hat E}{2 mc^2}  < \frac{k^2\hbar^2}{2m}\,, \quad
  \frac{\hbar q^2\hat E^2}{4k m^2c^3}  < \frac{k^2\hbar^2}{2m}
\end{equation}
have to be met.  The first two inequalities 
are both equivalent to
\begin{subequations}
  \begin{equation}
    \label{eq:perturbative_cond_a}
    \frac{|q|\hat E}{k^2\hbar c}<1
  \end{equation}
  and from the third inequality of \eqref{eq:perturbative_cond}
  follows with \eqref{eq:perturbative_cond_a}
  \begin{equation}
    \label{eq:perturbative_cond_b}
    \frac{|q|\hat E}{2k m c^2}<1\,.
  \end{equation}
\end{subequations}
For laser wavelengths that are longer than half of the Compton
wavelength the condition \eqref{eq:perturbative_cond_b} is always
fulfilled provided that the condition \eqref{eq:perturbative_cond_a} is
met.  In Fig.~\ref{fig:spin_dynamics_E} we show the spin dynamics for
three different peak electric field strengths, for the cases ${|q|\hat
  E}/({k^2\hbar c})<1$ with $\hat E=2.06\times10^{14}\,\mathrm{V/m}$,
${|q|\hat E}/({k^2\hbar c})\approx1$ with $\hat
E=3.09\times10^{14}\,\mathrm{V/m}$, and ${|q|\hat E}/({k^2\hbar c})>1$
with $\hat E=4.11\times10^{14}\,\mathrm{V/m}$, respectively.
The figure clearly shows how the inharmonicities start to set in at
${|q|\hat E}/({k^2\hbar c})\approx1$.

\begin{figure}[b]
  \centering
  \includegraphics[scale=0.5]{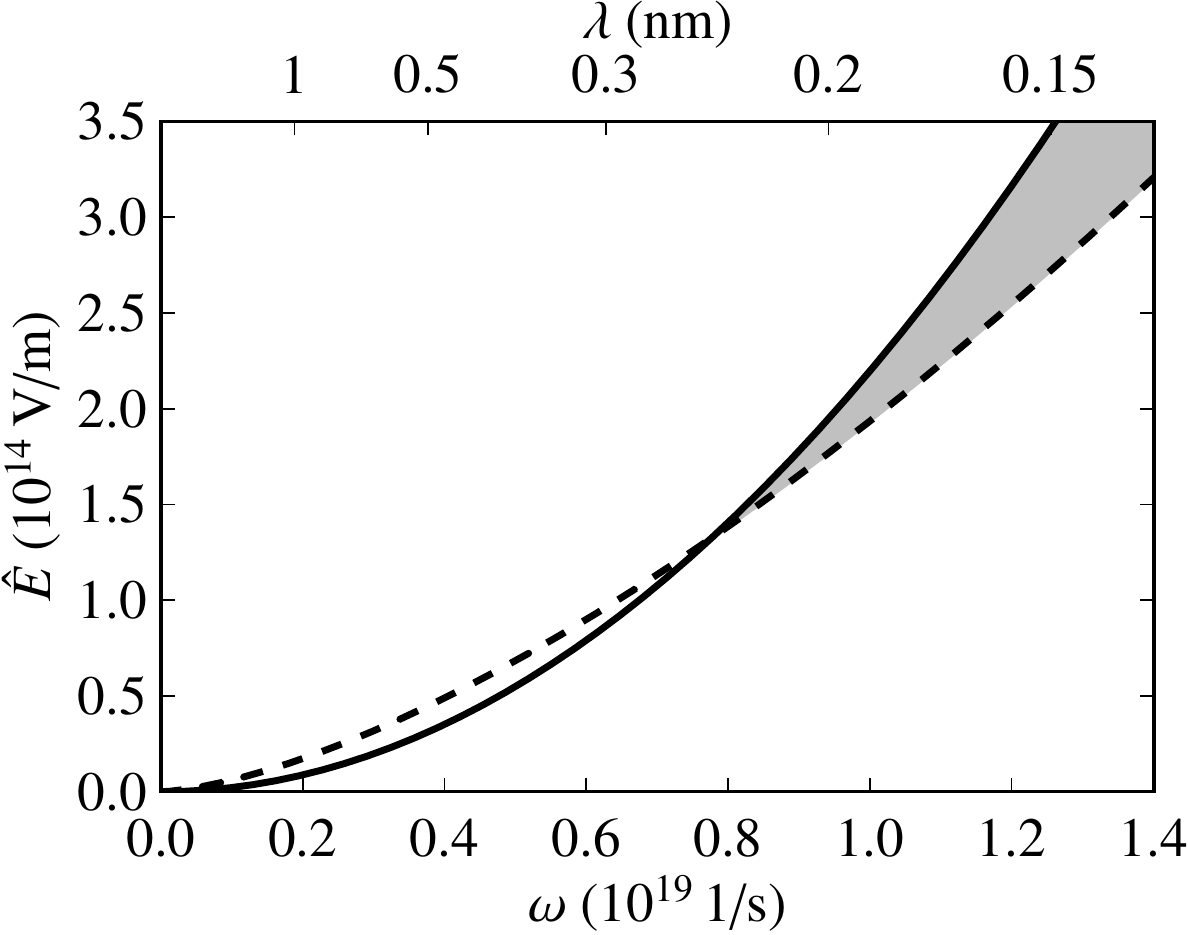}
  \caption{Laser parameters for which a full spin flip is feasible
    (gray shaded area). The upper and lower bounds are given by
    \eqref{eq:bounds_E} assuming that the electron can stay for
    $\mathcal{N}=5000$ laser cycles in the electromagnetic field's
    focus.}
  \label{fig:experimental_parameters}
\end{figure}

Generally speaking the observed electron-spin dynamics is not a
strong-field effect.  Spin flips may be observed also for weak laser
fields but the spin-precession frequency becomes small in this regime.
In this case, the electron must be trapped for many laser cycles in
the laser field's focus to observe a full spin flip.  If one assumes
that the electron stays for $\mathcal{N}$ cycles in the laser's focus,
then the required electric field strength is bounded as
\begin{equation}
  \label{eq:bounds_E}
  \left({\frac{\omega^6\hbar^2m^2}{2\mathcal{N}q^4}}\right)^{1/4} =
  \left({\frac{(2\pi)^6 c^6\hbar^2m^2}{2\mathcal{N}q^4\lambda^6}}\right)^{1/4} < 
  \hat E < 
  \frac{\omega^2\hbar}{|q|c}=
  \frac{(2\pi)^2c\hbar}{|q|\lambda^2}\,.
\end{equation}
The lower bound follows from $\mathcal{N}>\omega/(2\Omega)$ and the
upper bound from \eqref{eq:perturbative_cond_a}.  For the lower limit we
made the rather pessimistic assumption that a full spin flip would be
required to detect a change of the spin orientation.  More sensitive
electron-spin measurements would allow to employ shorter laser pulses
and/or lower intensities to verify the predicted effect.  Furthermore,
the inequalities in \eqref{eq:bounds_E} implicate that an increase of
the laser's wavelength may also require an increase of interaction time
in terms of laser cycles $\mathcal{N}$.  This is because when the
wavelength $\lambda$ is increased but all other parameters remain fixed,
then the lower bound will finally exceed the upper bound of
\eqref{eq:bounds_E} and then the effect becomes unverifiable.  Thus,
there is a specific range of wavelengths and electric field strengths
that render a full spin flip possible that depends on the parameter
$\mathcal{N}$ as illustrated in Fig.~\ref{fig:experimental_parameters}.

\section{Conclusions} 
\label{sec:Conclusions}

We investigated electron motion in two counterpropagating elliptically
polarized laser waves of opposite helicity and found spin precession
around the laser waves' propagation axis. The spin precession is caused
by the coupling of the light's spin density to the electron's spin
degree of freedom.  This mechanism is of similar origin as spin-orbit
coupling. The spin-precession frequency is directly proportional to the
product of the laser's intensity and the electromagnetic field's spin
density.  The observation of the predicted spin precession may become
feasible by employing near-future high-intensity x-ray laser facilities.
Furthermore, it might be interesting to generalize the setup considering
electron-vortex beams \cite{Hayrapetyan:etal:2013:Electron-Vortex_Beams}
and/or light-vortex beams and to study a possible interaction of the
light's spin or its orbital angular momentum with the electron beam's
orbital angular momentum.

\begin{acknowledgments}
  S.~A.\ acknowledges the nice hospitality during his visit at Illinois
  State University.  This work was supported by the NSF (USA).
\end{acknowledgments}

\bibliography{electron_spin_in_circular_light_long}

\begin{thebibliography}{31}%
\makeatletter
\providecommand \@ifxundefined [1]{%
 \@ifx{#1\undefined}
}%
\providecommand \@ifnum [1]{%
 \ifnum #1\expandafter \@firstoftwo
 \else \expandafter \@secondoftwo
 \fi
}%
\providecommand \@ifx [1]{%
 \ifx #1\expandafter \@firstoftwo
 \else \expandafter \@secondoftwo
 \fi
}%
\providecommand \natexlab [1]{#1}%
\providecommand \enquote  [1]{``#1''}%
\providecommand \bibnamefont  [1]{#1}%
\providecommand \bibfnamefont [1]{#1}%
\providecommand \citenamefont [1]{#1}%
\providecommand \href@noop [0]{\@secondoftwo}%
\providecommand \href [0]{\begingroup \@sanitize@url \@href}%
\providecommand \@href[1]{\@@startlink{#1}\@@href}%
\providecommand \@@href[1]{\endgroup#1\@@endlink}%
\providecommand \@sanitize@url [0]{\catcode `\\12\catcode `\$12\catcode
  `\&12\catcode `\#12\catcode `\^12\catcode `\_12\catcode `\%12\relax}%
\providecommand \@@startlink[1]{}%
\providecommand \@@endlink[0]{}%
\providecommand \url  [0]{\begingroup\@sanitize@url \@url }%
\providecommand \@url [1]{\endgroup\@href {#1}{\urlprefix }}%
\providecommand \urlprefix  [0]{URL }%
\providecommand \Eprint [0]{\href }%
\providecommand \doibase [0]{http://dx.doi.org/}%
\providecommand \selectlanguage [0]{\@gobble}%
\providecommand \bibinfo  [0]{\@secondoftwo}%
\providecommand \bibfield  [0]{\@secondoftwo}%
\providecommand \translation [1]{[#1]}%
\providecommand \BibitemOpen [0]{}%
\providecommand \bibitemStop [0]{}%
\providecommand \bibitemNoStop [0]{.\EOS\space}%
\providecommand \EOS [0]{\spacefactor3000\relax}%
\providecommand \BibitemShut  [1]{\csname bibitem#1\endcsname}%
\let\auto@bib@innerbib\@empty
\bibitem [{\citenamefont {Altarelli}\ \emph {et~al.}(2007)\citenamefont
  {Altarelli}, \citenamefont {Brinkmann}, \citenamefont {Chergui},
  \citenamefont {Decking}, \citenamefont {Dobson}, \citenamefont
  {D{\"u}sterer}, \citenamefont {Gr{\"u}bel}, \citenamefont {Graeff},
  \citenamefont {Graafsma}, \citenamefont {Hajdu}, \citenamefont {Marangos},
  \citenamefont {Pfl{\"u}ger}, \citenamefont {Redlin}, \citenamefont {Riley},
  \citenamefont {Robinson}, \citenamefont {Rossbach}, \citenamefont {Schwarz},
  \citenamefont {Tiedtke}, \citenamefont {Tschentscher}, \citenamefont
  {Vartaniants}, \citenamefont {Wabnitz}, \citenamefont {Weise}, \citenamefont
  {Wichmann}, \citenamefont {Witte}, \citenamefont {Wolf}, \citenamefont
  {Wulff},\ and\ \citenamefont {Yurkov}}]{Altarell:etal:2007:XFEL}%
  \BibitemOpen
  \bibinfo {editor} {\bibfnamefont {M.}~\bibnamefont {Altarelli}}, \bibinfo
  {editor} {\bibfnamefont {R.}~\bibnamefont {Brinkmann}}, \bibinfo {editor}
  {\bibfnamefont {M.}~\bibnamefont {Chergui}}, \bibinfo {editor} {\bibfnamefont
  {W.}~\bibnamefont {Decking}}, \bibinfo {editor} {\bibfnamefont
  {B.}~\bibnamefont {Dobson}}, \bibinfo {editor} {\bibfnamefont
  {S.}~\bibnamefont {D{\"u}sterer}}, \bibinfo {editor} {\bibfnamefont
  {G.}~\bibnamefont {Gr{\"u}bel}}, \bibinfo {editor} {\bibfnamefont
  {W.}~\bibnamefont {Graeff}}, \bibinfo {editor} {\bibfnamefont
  {H.}~\bibnamefont {Graafsma}}, \bibinfo {editor} {\bibfnamefont
  {J.}~\bibnamefont {Hajdu}} \emph {et~al.},\ eds.,\ \href@noop {} {\emph
  {\bibinfo {title} {The European X-Ray Free-Electron Laser Technical design
  report}}}\ (\bibinfo  {publisher} {DESY XFEL Project Group European XFEL
  Project Team Deutsches Elektronen-Synchrotron Member of the Helmholtz
  Association},\ \bibinfo {address} {Hamburg},\ \bibinfo {year}
  {2007})\BibitemShut {NoStop}%
\bibitem [{\citenamefont {Yanovsky}\ \emph {et~al.}(2008)\citenamefont
  {Yanovsky}, \citenamefont {Chvykov}, \citenamefont {Kalinchenko},
  \citenamefont {Rousseau}, \citenamefont {Planchon}, \citenamefont {Matsuoka},
  \citenamefont {Maksimchuk}, \citenamefont {Nees}, \citenamefont {Cheriaux},
  \citenamefont {Mourou},\ and\ \citenamefont
  {et~al.}}]{Yanovsky:Chvykov:etal:2008:Ultra-high_intensity-}%
  \BibitemOpen
  \bibfield  {author} {\bibinfo {author} {\bibfnamefont {V.}~\bibnamefont
  {Yanovsky}}, \bibinfo {author} {\bibfnamefont {V.}~\bibnamefont {Chvykov}},
  \bibinfo {author} {\bibfnamefont {G.}~\bibnamefont {Kalinchenko}}, \bibinfo
  {author} {\bibfnamefont {P.}~\bibnamefont {Rousseau}}, \bibinfo {author}
  {\bibfnamefont {T.}~\bibnamefont {Planchon}}, \bibinfo {author}
  {\bibfnamefont {T.}~\bibnamefont {Matsuoka}}, \bibinfo {author}
  {\bibfnamefont {A.}~\bibnamefont {Maksimchuk}}, \bibinfo {author}
  {\bibfnamefont {J.}~\bibnamefont {Nees}}, \bibinfo {author} {\bibfnamefont
  {G.}~\bibnamefont {Cheriaux}}, \bibinfo {author} {\bibfnamefont
  {G.}~\bibnamefont {Mourou}} \emph {et~al.},\ }\href {\doibase
  10.1364/OE.16.002109} {\bibfield  {journal} {\bibinfo  {journal} {Opt.
  Express}\ }\textbf {\bibinfo {volume} {16}},\ \bibinfo {pages} {2109}
  (\bibinfo {year} {2008})}\BibitemShut {NoStop}%
\bibitem [{\citenamefont {McNeil}\ and\ \citenamefont
  {Thompson}(2010)}]{McNeil:Thompson:2010:X-ray_free-electron}%
  \BibitemOpen
  \bibfield  {author} {\bibinfo {author} {\bibfnamefont {B.~W.~J.}\
  \bibnamefont {McNeil}}\ \bibnamefont {and}\ \bibinfo {author} {\bibfnamefont
  {N.~R.}\ \bibnamefont {Thompson}},\ }\href {\doibase
  10.1038/nphoton.2010.239} {\bibfield  {journal} {\bibinfo  {journal} {Nat.
  Photonics}\ }\textbf {\bibinfo {volume} {4}},\ \bibinfo {pages} {814}
  (\bibinfo {year} {2010})}\BibitemShut {NoStop}%
\bibitem [{\citenamefont {Emma}\ \emph {et~al.}(2010)\citenamefont {Emma},
  \citenamefont {Akre}, \citenamefont {Arthur}, \citenamefont {Bionta},
  \citenamefont {Bostedt}, \citenamefont {Bozek}, \citenamefont {Brachmann},
  \citenamefont {Bucksbaum}, \citenamefont {Coffee}, \citenamefont {Decker},
  \citenamefont {Ding}, \citenamefont {Dowell}, \citenamefont {Edstrom},
  \citenamefont {Fisher}, \citenamefont {Frisch}, \citenamefont {Gilevich},
  \citenamefont {Hastings}, \citenamefont {Hays}, \citenamefont {Hering},
  \citenamefont {Huang}, \citenamefont {Iverson}, \citenamefont {Loos},
  \citenamefont {Messerschmidt}, \citenamefont {Miahnahri}, \citenamefont
  {Moeller}, \citenamefont {Nuhn}, \citenamefont {Pile}, \citenamefont
  {Ratner}, \citenamefont {Rzepiela}, \citenamefont {Schultz}, \citenamefont
  {Smith}, \citenamefont {Stefan}, \citenamefont {Tompkins}, \citenamefont
  {Turner}, \citenamefont {Welch}, \citenamefont {White}, \citenamefont {Wu},
  \citenamefont {Yocky},\ and\ \citenamefont
  {Galayda}}]{Emma:Akre:etal:2010:First_lasing}%
  \BibitemOpen
  \bibfield  {author} {\bibinfo {author} {\bibfnamefont {P.}~\bibnamefont
  {Emma}}, \bibinfo {author} {\bibfnamefont {R.}~\bibnamefont {Akre}}, \bibinfo
  {author} {\bibfnamefont {J.}~\bibnamefont {Arthur}}, \bibinfo {author}
  {\bibfnamefont {R.}~\bibnamefont {Bionta}}, \bibinfo {author} {\bibfnamefont
  {C.}~\bibnamefont {Bostedt}}, \bibinfo {author} {\bibfnamefont
  {J.}~\bibnamefont {Bozek}}, \bibinfo {author} {\bibfnamefont
  {A.}~\bibnamefont {Brachmann}}, \bibinfo {author} {\bibfnamefont
  {P.}~\bibnamefont {Bucksbaum}}, \bibinfo {author} {\bibfnamefont
  {R.}~\bibnamefont {Coffee}}, \bibinfo {author} {\bibfnamefont {F.-J.}\
  \bibnamefont {Decker}} \emph {et~al.},\ }\href {\doibase
  10.1038/nphoton.2010.176} {\bibfield  {journal} {\bibinfo  {journal} {Nat.
  Photonics}\ }\textbf {\bibinfo {volume} {4}},\ \bibinfo {pages} {641}
  (\bibinfo {year} {2010})}\BibitemShut {NoStop}%
\bibitem [{\citenamefont {Mourou}\ and\ \citenamefont
  {Tajima}(2011)}]{Mourou:Tajima:2011:More_Intense}%
  \BibitemOpen
  \bibfield  {author} {\bibinfo {author} {\bibfnamefont {G.}~\bibnamefont
  {Mourou}}\ \bibnamefont {and}\ \bibinfo {author} {\bibfnamefont
  {T.}~\bibnamefont {Tajima}},\ }\href {\doibase 10.1126/science.1200292}
  {\bibfield  {journal} {\bibinfo  {journal} {Science}\ }\textbf {\bibinfo
  {volume} {331}},\ \bibinfo {pages} {41} (\bibinfo {year} {2011})}\BibitemShut
  {NoStop}%
\bibitem [{\citenamefont {Mourou}\ \emph {et~al.}(2012)\citenamefont {Mourou},
  \citenamefont {Fisch}, \citenamefont {Malkin}, \citenamefont {Toroker},
  \citenamefont {Khazanov}, \citenamefont {Sergeev}, \citenamefont {Tajima},\
  and\ \citenamefont
  {Le~Garrec}}]{Mourou:Fisch:etal:2012:Exawatt-Zettawatt_pulse}%
  \BibitemOpen
  \bibfield  {author} {\bibinfo {author} {\bibfnamefont {G.}~\bibnamefont
  {Mourou}}, \bibinfo {author} {\bibfnamefont {N.}~\bibnamefont {Fisch}},
  \bibinfo {author} {\bibfnamefont {V.}~\bibnamefont {Malkin}}, \bibinfo
  {author} {\bibfnamefont {Z.}~\bibnamefont {Toroker}}, \bibinfo {author}
  {\bibfnamefont {E.}~\bibnamefont {Khazanov}}, \bibinfo {author}
  {\bibfnamefont {A.}~\bibnamefont {Sergeev}}, \bibinfo {author} {\bibfnamefont
  {T.}~\bibnamefont {Tajima}},\ \bibnamefont {and}\ \bibinfo {author}
  {\bibfnamefont {B.}~\bibnamefont {Le~Garrec}},\ }\href {\doibase
  10.1016/j.optcom.2011.10.089} {\bibfield  {journal} {\bibinfo  {journal}
  {Opt. Commun.}\ }\textbf {\bibinfo {volume} {285}},\ \bibinfo {pages} {720}
  (\bibinfo {year} {2012})}\BibitemShut {NoStop}%
\bibitem [{\citenamefont
  {Brabec}(2008)}]{Brabec:2008:Strong_Field_Laser_Physics}%
  \BibitemOpen
  \bibinfo {editor} {\bibfnamefont {T.}~\bibnamefont {Brabec}},\ ed.,\
  \href@noop {} {\emph {\bibinfo {title} {Strong Field Laser Physics}}},\
  \bibinfo {series} {Springer Series in Optical Sciences}, Vol.\ \bibinfo
  {volume} {134}\ (\bibinfo  {publisher} {Springer},\ \bibinfo {address}
  {Heidelberg},\ \bibinfo {year} {2008})\BibitemShut {NoStop}%
\bibitem [{\citenamefont {Ehlotzky}\ \emph {et~al.}(2009)\citenamefont
  {Ehlotzky}, \citenamefont {Krajewska},\ and\ \citenamefont
  {Kami\'nski}}]{Ehlotzky:Krajewska:Kaminski:2009:Fundamental_processes}%
  \BibitemOpen
  \bibfield  {author} {\bibinfo {author} {\bibfnamefont {F.}~\bibnamefont
  {Ehlotzky}}, \bibinfo {author} {\bibfnamefont {K.}~\bibnamefont
  {Krajewska}},\ \bibnamefont {and}\ \bibinfo {author} {\bibfnamefont {J.~Z.}\
  \bibnamefont {Kami\'nski}},\ }\href {\doibase 10.1088/0034-4885/72/4/046401}
  {\bibfield  {journal} {\bibinfo  {journal} {Rep. Prog. Phys.}\ }\textbf
  {\bibinfo {volume} {72}},\ \bibinfo {pages} {046401} (\bibinfo {year}
  {2009})}\BibitemShut {NoStop}%
\bibitem [{\citenamefont {Di~Piazza}\ \emph {et~al.}(2012)\citenamefont
  {Di~Piazza}, \citenamefont {M\"uller}, \citenamefont {Hatsagortsyan},\ and\
  \citenamefont
  {Keitel}}]{Di-Piazza:Muller:etal:2012:Extremely_high-intensity}%
  \BibitemOpen
  \bibfield  {author} {\bibinfo {author} {\bibfnamefont {A.}~\bibnamefont
  {Di~Piazza}}, \bibinfo {author} {\bibfnamefont {C.}~\bibnamefont {M\"uller}},
  \bibinfo {author} {\bibfnamefont {K.~Z.}\ \bibnamefont {Hatsagortsyan}},\
  \bibnamefont {and}\ \bibinfo {author} {\bibfnamefont {C.~H.}\ \bibnamefont
  {Keitel}},\ }\href {\doibase 10.1103/RevModPhys.84.1177} {\bibfield
  {journal} {\bibinfo  {journal} {Rev. Mod. Phys.}\ }\textbf {\bibinfo {volume}
  {84}},\ \bibinfo {pages} {1177} (\bibinfo {year} {2012})}\BibitemShut
  {NoStop}%
\bibitem [{\citenamefont {Walser}\ \emph {et~al.}(2002)\citenamefont {Walser},
  \citenamefont {Urbach}, \citenamefont {Hatsagortsyan}, \citenamefont {Hu},\
  and\ \citenamefont {Keitel}}]{Walser:Urbach:etal:2002:Spin_and}%
  \BibitemOpen
  \bibfield  {author} {\bibinfo {author} {\bibfnamefont {M.~W.}\ \bibnamefont
  {Walser}}, \bibinfo {author} {\bibfnamefont {D.~J.}\ \bibnamefont {Urbach}},
  \bibinfo {author} {\bibfnamefont {K.~Z.}\ \bibnamefont {Hatsagortsyan}},
  \bibinfo {author} {\bibfnamefont {S.~X.}\ \bibnamefont {Hu}},\ \bibnamefont
  {and}\ \bibinfo {author} {\bibfnamefont {C.~H.}\ \bibnamefont {Keitel}},\
  }\href {\doibase 10.1103/PhysRevA.65.043410} {\bibfield  {journal} {\bibinfo
  {journal} {Phys. Rev. A}\ }\textbf {\bibinfo {volume} {65}},\ \bibinfo
  {pages} {043410} (\bibinfo {year} {2002})}\BibitemShut {NoStop}%
\bibitem [{\citenamefont {Ahrens}\ \emph {et~al.}(2012)\citenamefont {Ahrens},
  \citenamefont {Bauke}, \citenamefont {Keitel},\ and\ \citenamefont
  {M\"uller}}]{Ahrens:Bauke:etal:2012:Spin_Dynamics}%
  \BibitemOpen
  \bibfield  {author} {\bibinfo {author} {\bibfnamefont {S.}~\bibnamefont
  {Ahrens}}, \bibinfo {author} {\bibfnamefont {H.}~\bibnamefont {Bauke}},
  \bibinfo {author} {\bibfnamefont {C.~H.}\ \bibnamefont {Keitel}},\
  \bibnamefont {and}\ \bibinfo {author} {\bibfnamefont {C.}~\bibnamefont
  {M\"uller}},\ }\href {\doibase 10.1103/PhysRevLett.109.043601} {\bibfield
  {journal} {\bibinfo  {journal} {Phys. Rev. Lett.}\ }\textbf {\bibinfo
  {volume} {109}},\ \bibinfo {pages} {043601} (\bibinfo {year}
  {2012})}\BibitemShut {NoStop}%
\bibitem [{\citenamefont {Ahrens}\ \emph {et~al.}(2013)\citenamefont {Ahrens},
  \citenamefont {Bauke}, \citenamefont {Keitel},\ and\ \citenamefont
  {M\"uller}}]{Ahrens:Bauke:etal:2013:Kapitza-Dirac_effect}%
  \BibitemOpen
  \bibfield  {author} {\bibinfo {author} {\bibfnamefont {S.}~\bibnamefont
  {Ahrens}}, \bibinfo {author} {\bibfnamefont {H.}~\bibnamefont {Bauke}},
  \bibinfo {author} {\bibfnamefont {C.~H.}\ \bibnamefont {Keitel}},\
  \bibnamefont {and}\ \bibinfo {author} {\bibfnamefont {C.}~\bibnamefont
  {M\"uller}},\ }\href {\doibase 10.1103/PhysRevA.88.012115} {\bibfield
  {journal} {\bibinfo  {journal} {Phys. Rev. A}\ }\textbf {\bibinfo {volume}
  {88}},\ \bibinfo {pages} {012115} (\bibinfo {year} {2013})}\BibitemShut
  {NoStop}%
\bibitem [{\citenamefont {Bauke}\ \emph {et~al.}(2014)\citenamefont {Bauke},
  \citenamefont {Ahrens}, \citenamefont {Keitel},\ and\ \citenamefont
  {Grobe}}]{Bauke:Ahrens:etal:2014:Electron-spin_dynamics}%
  \BibitemOpen
  \bibfield  {author} {\bibinfo {author} {\bibfnamefont {H.}~\bibnamefont
  {Bauke}}, \bibinfo {author} {\bibfnamefont {S.}~\bibnamefont {Ahrens}},
  \bibinfo {author} {\bibfnamefont {C.~H.}\ \bibnamefont {Keitel}},\
  \bibnamefont {and}\ \bibinfo {author} {\bibfnamefont {R.}~\bibnamefont
  {Grobe}},\ }\href {\doibase 10.1088/1367-2630/16/10/103028} {\bibfield
  {journal} {\bibinfo  {journal} {New J. Phys.}\ }\textbf {\bibinfo {volume}
  {16}},\ \bibinfo {pages} {103028} (\bibinfo {year} {2014})}\BibitemShut
  {NoStop}%
\bibitem [{\citenamefont {Zhang}\ and\ \citenamefont
  {Niu}(2014)}]{Zhang:Niu:2014:Angular_Momentum}%
  \BibitemOpen
  \bibfield  {author} {\bibinfo {author} {\bibfnamefont {L.}~\bibnamefont
  {Zhang}}\ \bibnamefont {and}\ \bibinfo {author} {\bibfnamefont
  {Q.}~\bibnamefont {Niu}},\ }\href {\doibase 10.1103/physrevlett.112.085503}
  {\bibfield  {journal} {\bibinfo  {journal} {Phys. Rev. Lett.}\ }\textbf
  {\bibinfo {volume} {112}},\ \bibinfo {pages} {085503} (\bibinfo {year}
  {2014})}\BibitemShut {NoStop}%
\bibitem [{\citenamefont {Mondal}\ \emph {et~al.}(2014)\citenamefont {Mondal},
  \citenamefont {Deb},\ and\ \citenamefont
  {Majumder}}]{Mondal:etal:2014:Angular_Momentum_Transfer}%
  \BibitemOpen
  \bibfield  {author} {\bibinfo {author} {\bibfnamefont {P.~K.}\ \bibnamefont
  {Mondal}}, \bibinfo {author} {\bibfnamefont {B.}~\bibnamefont {Deb}},\
  \bibnamefont {and}\ \bibinfo {author} {\bibfnamefont {S.}~\bibnamefont
  {Majumder}},\ }\href {\doibase 10.1103/PhysRevA.89.063418} {\bibfield
  {journal} {\bibinfo  {journal} {Phys. Rev. A}\ }\textbf {\bibinfo {volume}
  {89}},\ \bibinfo {pages} {063418} (\bibinfo {year} {2014})}\BibitemShut
  {NoStop}%
\bibitem [{\citenamefont {Beth}(1936)}]{Beth:1936:Mechanical_Detection}%
  \BibitemOpen
  \bibfield  {author} {\bibinfo {author} {\bibfnamefont {R.}~\bibnamefont
  {Beth}},\ }\href {\doibase 10.1103/PhysRev.50.115} {\bibfield  {journal}
  {\bibinfo  {journal} {Phys. Rev.}\ }\textbf {\bibinfo {volume} {50}},\
  \bibinfo {pages} {115} (\bibinfo {year} {1936})}\BibitemShut {NoStop}%
\bibitem [{\citenamefont
  {Jackson}(1998)}]{Jackson:1998:Classical_Electrodynamics}%
  \BibitemOpen
  \bibfield  {author} {\bibinfo {author} {\bibfnamefont {J.~D.}\ \bibnamefont
  {Jackson}},\ }\href@noop {} {\emph {\bibinfo {title} {Classical
  Electrodynamics}}}\ (\bibinfo  {publisher} {John Wiley \& Sons},\ \bibinfo
  {address} {New York},\ \bibinfo {year} {1998})\BibitemShut {NoStop}%
\bibitem [{\citenamefont {Stewart}(2005)}]{Stewart:2005:Angular_momentum}%
  \BibitemOpen
  \bibfield  {author} {\bibinfo {author} {\bibfnamefont {A.~M.}\ \bibnamefont
  {Stewart}},\ }\href {\doibase 10.1080/09500340512331326832} {\bibfield
  {journal} {\bibinfo  {journal} {J. Mod. Opt.}\ }\textbf {\bibinfo {volume}
  {52}},\ \bibinfo {pages} {1145} (\bibinfo {year} {2005})}\BibitemShut
  {NoStop}%
\bibitem [{\citenamefont {Barnett}(2011)}]{Barnett:2011:On_six}%
  \BibitemOpen
  \bibfield  {author} {\bibinfo {author} {\bibfnamefont {S.~M.}\ \bibnamefont
  {Barnett}},\ }\href {\doibase 10.1088/2040-8978/13/6/064010} {\bibfield
  {journal} {\bibinfo  {journal} {J. Opt. (Bristol, U. K.)}\ }\textbf {\bibinfo
  {volume} {13}},\ \bibinfo {pages} {064010} (\bibinfo {year}
  {2011})}\BibitemShut {NoStop}%
\bibitem [{\citenamefont {Cameron}\ and\ \citenamefont
  {Barnett}(2012)}]{Cameron:Barnett:2012:Electric-magnetic_symmetry}%
  \BibitemOpen
  \bibfield  {author} {\bibinfo {author} {\bibfnamefont {R.~P.}\ \bibnamefont
  {Cameron}}\ \bibnamefont {and}\ \bibinfo {author} {\bibfnamefont {S.~M.}\
  \bibnamefont {Barnett}},\ }\href {\doibase 10.1088/1367-2630/14/12/123019}
  {\bibfield  {journal} {\bibinfo  {journal} {New J. Phys.}\ }\textbf {\bibinfo
  {volume} {14}},\ \bibinfo {pages} {123019} (\bibinfo {year}
  {2012})}\BibitemShut {NoStop}%
\bibitem [{\citenamefont {Bliokh}\ \emph {et~al.}(2013)\citenamefont {Bliokh},
  \citenamefont {Bekshaev},\ and\ \citenamefont
  {Nori}}]{Bliokh:Bekshaev:etal:2013:Dual_electromagnetism}%
  \BibitemOpen
  \bibfield  {author} {\bibinfo {author} {\bibfnamefont {K.~Y.}\ \bibnamefont
  {Bliokh}}, \bibinfo {author} {\bibfnamefont {A.~Y.}\ \bibnamefont
  {Bekshaev}},\ \bibnamefont {and}\ \bibinfo {author} {\bibfnamefont
  {F.}~\bibnamefont {Nori}},\ }\href {\doibase 10.1088/1367-2630/15/3/033026}
  {\bibfield  {journal} {\bibinfo  {journal} {New J. Phys.}\ }\textbf {\bibinfo
  {volume} {15}},\ \bibinfo {pages} {033026} (\bibinfo {year}
  {2013})}\BibitemShut {NoStop}%
\bibitem [{\citenamefont {Barnett}(2010)}]{Barnett:2010:Rotation_of}%
  \BibitemOpen
  \bibfield  {author} {\bibinfo {author} {\bibfnamefont {S.~M.}\ \bibnamefont
  {Barnett}},\ }\href {\doibase 10.1080/09500341003654427} {\bibfield
  {journal} {\bibinfo  {journal} {Journal of Modern Optics}\ }\textbf {\bibinfo
  {volume} {57}},\ \bibinfo {pages} {1339} (\bibinfo {year}
  {2010})}\BibitemShut {NoStop}%
\bibitem [{\citenamefont {Cameron}(2014)}]{Cameron:2014:On_second}%
  \BibitemOpen
  \bibfield  {author} {\bibinfo {author} {\bibfnamefont {R.~P.}\ \bibnamefont
  {Cameron}},\ }\href {\doibase 10.1088/2040-8978/16/1/015708} {\bibfield
  {journal} {\bibinfo  {journal} {J. Opt.}\ }\textbf {\bibinfo {volume} {16}},\
  \bibinfo {pages} {015708} (\bibinfo {year} {2014})}\BibitemShut {NoStop}%
\bibitem [{\citenamefont {Gross}(2004)}]{Gross:2004:Relativistic_Quantum}%
  \BibitemOpen
  \bibfield  {author} {\bibinfo {author} {\bibfnamefont {F.}~\bibnamefont
  {Gross}},\ }\href@noop {} {\emph {\bibinfo {title} {Relativistic Quantum
  Mechanics and Field Theory}}}\ (\bibinfo  {publisher} {Wiley-VCH},\ \bibinfo
  {address} {Weinheim},\ \bibinfo {year} {2004})\BibitemShut {NoStop}%
\bibitem [{\citenamefont {Thaller}(2005)}]{Thaller:2005:Advanced_Visual}%
  \BibitemOpen
  \bibfield  {author} {\bibinfo {author} {\bibfnamefont {B.}~\bibnamefont
  {Thaller}},\ }\href@noop {} {\emph {\bibinfo {title} {Advanced Visual Quantum
  Mechanics}}}\ (\bibinfo  {publisher} {Springer},\ \bibinfo {address}
  {Heidelberg},\ \bibinfo {year} {2005})\BibitemShut {NoStop}%
\bibitem [{\citenamefont {Foldy}\ and\ \citenamefont
  {Wouthuysen}(1950)}]{Foldy:Wouthuysen:1950:Non-Relativistic_Limit}%
  \BibitemOpen
  \bibfield  {author} {\bibinfo {author} {\bibfnamefont {L.~L.}\ \bibnamefont
  {Foldy}}\ \bibnamefont {and}\ \bibinfo {author} {\bibfnamefont {S.~A.}\
  \bibnamefont {Wouthuysen}},\ }\href {\doibase 10.1103/PhysRev.78.29}
  {\bibfield  {journal} {\bibinfo  {journal} {Phys. Rev.}\ }\textbf {\bibinfo
  {volume} {78}},\ \bibinfo {pages} {29} (\bibinfo {year} {1950})}\BibitemShut
  {NoStop}%
\bibitem [{\citenamefont
  {de~Vries}(1970)}]{Vries:1970:Foldy-Wouthuysen_Transformations}%
  \BibitemOpen
  \bibfield  {author} {\bibinfo {author} {\bibfnamefont {E.}~\bibnamefont
  {de~Vries}},\ }\href {\doibase 10.1002/prop.19700180402} {\bibfield
  {journal} {\bibinfo  {journal} {Fortschr. Phys.}\ }\textbf {\bibinfo {volume}
  {18}},\ \bibinfo {pages} {149} (\bibinfo {year} {1970})}\BibitemShut
  {NoStop}%
\bibitem [{\citenamefont {Fr\"ohlich}\ and\ \citenamefont
  {Studer}(1993)}]{Frohlich:Studer:1993:Gauge_invariance}%
  \BibitemOpen
  \bibfield  {author} {\bibinfo {author} {\bibfnamefont {J.}~\bibnamefont
  {Fr\"ohlich}}\ \bibnamefont {and}\ \bibinfo {author} {\bibfnamefont
  {U.}~\bibnamefont {Studer}},\ }\href {\doibase 10.1103/revmodphys.65.733}
  {\bibfield  {journal} {\bibinfo  {journal} {Rev. Mod. Phys.}\ }\textbf
  {\bibinfo {volume} {65}},\ \bibinfo {pages} {733} (\bibinfo {year}
  {1993})}\BibitemShut {NoStop}%
\bibitem [{\citenamefont {Joachain}\ \emph {et~al.}(2011)\citenamefont
  {Joachain}, \citenamefont {Kylstra},\ and\ \citenamefont
  {Potvliege}}]{Joachain:etal:2011:Atoms}%
  \BibitemOpen
  \bibfield  {author} {\bibinfo {author} {\bibfnamefont {C.~J.}\ \bibnamefont
  {Joachain}}, \bibinfo {author} {\bibfnamefont {N.~J.}\ \bibnamefont
  {Kylstra}},\ \bibnamefont {and}\ \bibinfo {author} {\bibfnamefont {R.~M.}\
  \bibnamefont {Potvliege}},\ }\href {\doibase 10.1017/CBO9780511993459} {\emph
  {\bibinfo {title} {Atoms in Intense Laser Fields}}}\ (\bibinfo  {publisher}
  {Cambridge University Press},\ \bibinfo {address} {Cambridge},\ \bibinfo
  {year} {2011})\BibitemShut {NoStop}%
\bibitem [{\citenamefont {Ahrens}(2012)}]{Ahrens:2012:PhD_thesis}%
  \BibitemOpen
  \bibfield  {author} {\bibinfo {author} {\bibfnamefont {S.}~\bibnamefont
  {Ahrens}},\ }\emph {\bibinfo {title} {Investigation of the {K}apitza-{D}irac
  effect in the relativistic regime}},\ \href@noop {} {Ph.D. thesis},\ \bibinfo
   {school} {{Ruprecht-Karls Universit\"at Heidelberg}} (\bibinfo {year}
  {2012}),\ \bibinfo {note}
  {\url{http://www.ub.uni-heidelberg.de/archiv/14049}}\BibitemShut {NoStop}%
\bibitem [{\citenamefont {Hayrapetyan}\ \emph {et~al.}(2014)\citenamefont
  {Hayrapetyan}, \citenamefont {Matula}, \citenamefont {Aiello}, \citenamefont
  {Surzhykov},\ and\ \citenamefont
  {Fritzsche}}]{Hayrapetyan:etal:2013:Electron-Vortex_Beams}%
  \BibitemOpen
  \bibfield  {author} {\bibinfo {author} {\bibfnamefont {A.~G.}\ \bibnamefont
  {Hayrapetyan}}, \bibinfo {author} {\bibfnamefont {O.}~\bibnamefont {Matula}},
  \bibinfo {author} {\bibfnamefont {A.}~\bibnamefont {Aiello}}, \bibinfo
  {author} {\bibfnamefont {A.}~\bibnamefont {Surzhykov}},\ \bibnamefont {and}\
  \bibinfo {author} {\bibfnamefont {S.}~\bibnamefont {Fritzsche}},\ }\href
  {\doibase 10.1103/PhysRevLett.112.134801} {\bibfield  {journal} {\bibinfo
  {journal} {Phys. Rev. Lett.}\ }\textbf {\bibinfo {volume} {112}},\ \bibinfo
  {pages} {134801} (\bibinfo {year} {2014})}\BibitemShut {NoStop}%
\end{thebibliography}%

\end{document}